\documentclass{article}
\usepackage[margin=2.5cm]{geometry}
\usepackage[utf8]{inputenc}
\usepackage{amssymb, amsmath}

\usepackage{enumitem}
\usepackage{booktabs}

\usepackage{url, xcolor}

\usepackage{graphicx}
\usepackage{settings_custom}
\usepackage[english]{babel}

\usepackage{graphicx}
\usepackage{caption}
\usepackage{subcaption}
\usepackage[colorlinks=true, allcolors=blue]{hyperref}
\usepackage{booktabs}
\usepackage{enumerate}
\usepackage{bbm}
\usepackage[toc,page]{appendix}
\graphicspath{{}}
\usepackage{float}
\usepackage{graphicx}

\title{Dynamical Cavity Method for Hypergraphs and its \\ Application to Quenches in the \texorpdfstring{$\boldsymbol{k}$}{}-XOR-SAT Problem}
\author{Aude Maier, Freya Behrens, and Lenka Zdeborová}
\date{\footnotesize
    Statistical Physics of Computation Laboratory,\\
    École polytechnique fédérale de Lausanne (EPFL)
    CH-1015 Lausanne
}

\begin{document}
    \maketitle

\newcommand{\fb}[1]{{\color{magenta} [FB: #1]}}
\newcommand{\am}[1]{{\color{blue} [AM: #1]}}
\newcommand{\outerF}{\mathrm{outer}}
\newcommand{\innerF}{\mathrm{inner}}
\newcommand{\attr}{\mathrm{attr}}
\newcommand{\initF}{\mathrm{init}}
\newcommand{\RS}{\mathrm{RS}}
\begin{abstract}
The dynamical cavity method and its backtracking version provide a powerful approach to studying the properties of dynamical processes on large random graphs.
This paper extends these methods to hypergraphs, enabling the analysis of interactions involving more than two variables.
We apply them to analyse the $k$-XOR-satisfiability ($k$-XOR-SAT) problem, an important model in theoretical computer science which is closely related to the diluted $p$-spin model from statistical physics.
In particular, we examine whether the quench dynamics -- a deterministic, locally greedy process -- can find solutions with only a few violated constraints on $d$-regular $k$-uniform hypergraphs.
Our results demonstrate that the methods accurately characterize the attractors of the dynamics. It enables us to compute the energy reached by typical trajectories of the dynamical process in different parameter regimes.
We show that these predictions are accurate, including cases where a classical mean-field approach fails.
\end{abstract}

\maketitle

\section{Introduction}
The past decades have seen an explosion of applications of methods from statistical physics to study constraint satisfaction and optimization problems.
In particular, the cavity method has shown to be remarkably well-suited to find equilibrium properties in the thermodynamic limit for systems from theoretical computer science ~\cite{altarelli2008review,mézard2009information}.
However, many questions relevant to optimization problems concern the performance of different algorithms -- the dynamical processes on the graph, rather than the static properties of its energy landscape.

The cavity method conducts only a \textit{static} analysis on the existence of solutions for a particular system.
While this analysis has some implications for a specific group of algorithms~\cite{Gamarnik_2021}, it does not concern other classes of simple search algorithms.
The absence of exact solutions to this type of question led to the development of methods that take full trajectories of the dynamics account, namely the \textit{dynamical cavity method} (DCM)~\cite{Neri_2009-DCM,Lokhov_2015-DCM,Mimura_2009-DCM,Kanoria_2011-DCM,Hatchett_2004-DCM}, and more recently its \textit{backtracking} version (BCDM) ~\cite{BDCM}.

In the context of these developments, the first key contribution of this work is to present a generalization of both methods to the study of processes involving interactions between more than two variables, i.e. dynamical processes on hypergraphs.
This is useful, because hypergraphs provide a general framework to represent constraint satisfaction problems with a wide range of applications including scheduling and location problems~\cite{Carraresi_1993-hypergraphs,Bretto2013-hypergraphs,hypergraphs,article-hypergraphs}.

As a direct application of this generalization, we consider the sparse $k$-XOR-satisfiability~($k$-XOR-SAT) problem,  a variant of the classical Boolean satisfiability problem~($k$-SAT)~\cite{Ibrahimi_2015-sat}.
In $k$-XOR-SAT, boolean variables are subject to constraints represented by XOR (exclusive OR) relations each involving $k$ variables.
This is closely related to the diluted $p$-spin model in statistical physics.
For this reason, the problem gathered great interest in the field of disordered systems~\cite{mezard2002alternative,Cocco_2003-glass,Franz_2001-glass,Ricci_Tersenghi_2001-glass}. 

Although the case $p=2$ of the diluted $p$-spin model already offers a model with interesting properties (e.g. as the Ising or Sherrington-Kirkpatrick model), models with $p>2$ exhibit a more complex and rugged energy landscape. In particular, these are the prototypical models exhibiting the random-first-order type of phase transition with dynamical and Kauzmann transitions that were used to study the properties of more general glassy systems~\cite{xia2000fragilities,biroli2012random}.

While the evolution of simple algorithms was already analyzed for the $p=2$ case in~\cite{BDCM,BehrensCellularAutomata}, in this work, we extend this analysis to the richer landscape of the $k$-XOR-SAT (diluted $p$-spin) model.
More concretely, using the DCM and the BDCM, we seek to understand the convergence behavior of quench dynamics on $k$-XOR-SAT instances and evaluate their performance in minimizing the number of unsatisfied constraints.
In our context, a quench corresponds to a deterministic, synchronous and locally greedy evolution of the system from a highly entropic state to a low-energy out-of-equilibrium configuration.
We use this dynamical process as a simple search algorithm and we compare its performance to the equilibrium properties of the system, i.e. the satisfiability threshold and spin glass phase transition~\cite{mezard2002alternative,PITTEL_SORKIN_2016,GARDNER1985747}.

In the remainder of this work, we first develop the notation of general dynamical processes on hypergraphs in Section~\ref{sec:notation}, and for the $k$-XOR-SAT problem in particular in Section~\ref{sec:XOR-SAT}. The dynamical cavity and its backtracking generalization to hypergraphs are introduced in Section~\ref{sec:BDCM}, along with refinements for the $k$-XOR-SAT.
Together with a comparison to a naïve mean-field approach, we apply these methods to the $k$-XOR-SAT and discuss our findings in Section~\ref{sec:results}.

\section{Analysing Dynamical Processes on Hypergraphs}\label{sec:notation}

\subsection{Notation}
\paragraph{Hypergraphs.}
We define an \textit{undirected hypergraph} as a pair $G=(V,E)$ where $V=\{1,\ldots,n\}$ are the \textit{nodes} of the hypergraph and the \textit{hyperedges} $E=\{a,b,\ldots\}$ each contain a subsets of nodes, $a,b,\ldots \subseteq V$.
There are $|E| = m$ hyperedges.
When the context allows, we denote nodes using the letters $i,j,k$ and hyperedges with $a,b,c$.
The \textit{degree} of a node $i$, denoted $d_i$, is the number of hyperedges that contain it, $d_i=|\{a\in E\;|\;i\in a\}|$.
Similarly, the \textit{order} $k_a$ of a hyperedge $a$ is the number of nodes it contains, $k_a=|a|$.
The \textit{neighborhood} $\partial i$ of a node $i$ is the set of hyperedges that contain $i$, formally $\partial i=\{a\in E\;|\;i\in a\}$.
In this work we focus on regular and uniform hypergraphs: In a \textit{$d$-regular hypergraph} all nodes have the same degree $d$ and in a \textit{$k$-uniform hypergraph} all hyperedges have the same order $k$.
We note that a 2-uniform hypergraph is an ordinary graph, where hyperedges are called edges and each edge is a link between two nodes. 

\paragraph{Dynamics.}
In the case of a time discrete dynamical process on a $d$-regular and $k$-uniform hypergraph, each node is assigned, at each time step from time $t=0$ until $T$, one of the discrete states from a finite set $S$.
We denote the state of node $i$ as $x_i \in S$ and the states of all nodes belonging to a subset $a\subseteq V$ as $\mathbf{x}_a=\{x_i\;|\;\forall i \in a\}$.
The state of the complete hypergraph, the \textit{configuration}, is given by the vector $\mathbf{x}=(x_1,\ldots,x_n)\in S^n$.

The dynamical process is described by a map $F:S^n\rightarrow S^n$, the \textit{global update rule}, that gives the configuration at the next time step as a function of the current configuration, $\mathbf{x}^{(t+1)}=F(\mathbf{x}^{(t)})$.
Under the assumption that the next state of a node only depends on its current state and the states of the nodes contained in its neighborhood, $F$ can be written as a set of \textit{local update rules} $f_i:S\times S^{d\cdot (k-1)}\rightarrow S$ such that $x_i^{(t+1)} = f_i(x_i^{(t)};\{\mathbf{x}_{a\backslash i}^{(t)}\}_{a\in\partial i})$.
Note that in this process, the update of the nodes is synchronous, deterministic and independent of the order of the neighbors.

Since a hyperedge represents an interaction between $k$ nodes, in this paper we decompose the local update rule $f_i$ as:
\begin{equation}
    x_i^{(t+1)} = f_i^{\outerF}\left(x_i^{(t)};\{f_i^{\innerF}(x_i^{(t)};\mathbf{x}_{a\backslash i}^{(t)})\}_{a\in\partial i}\right)\,,
\end{equation}
where the output of $f_i^{\innerF}:S\times S^{k-1}\rightarrow S$ can be seen as the preference of hyperedge $a$ for the next value taken by node $i$, and $f_i^{\outerF}:S\times S^{d}\rightarrow S$ aggregates all received inputs from its edges to update $x_i$.
This representation allows us to treat the influence of a node $j$ on the update of $i$ as part of the preference of the hyperedge $a$. 
The introduction of the inner local update rule $f_i^{\innerF}:S\times S^{k-1}\rightarrow S$ removes part of the information accessible to the node $i$, as the individual values of the nodes in the neighboring hyperedges are no more known.
It restricts the space of update rules that can be written in this framework.

A \textit{trajectory of length t} is defined as a trajectory of $t$ configurations $\underline{\mathbf{x}}:=(\mathbf{x}^{(1)}, \ldots, \mathbf{x}^{(t)})$ with the additional constraint that $F(\mathbf{x}^{(\tau)})=\mathbf{x}^{(\tau+1)}$ for all $\tau=1,\ldots,t-1$.
Additionally, a sequence $(\mathbf{x}^{(1)}, \ldots, \mathbf{x}^{(c)})$ such that $F(\mathbf{x}^{(c)})=\mathbf{x}^{(1)}$ is called a \textit{limit cycle} or \textit{attractor of length c}.
Since the configuration space $S^n$ is finite and the update rule $F$ is deterministic, the discrete temporal evolution of any initial configuration $\mathbf{x}^{(1)}$ is guaranteed to eventually reach either a stable configuration $\mathbf{x}=F(\mathbf{x})$, or a limit cycle $(\mathbf{x}^{(1)}, \ldots, \mathbf{x}^{(c)})$.
Note that a stable configuration is a limit cycle of length $c=1$.
When considering a trajectory $\underline{\mathbf{x}}:=(\mathbf{x}^{(1)}, \ldots, \mathbf{x}^{(t)}, \mathbf{x}^{(t+1)}, \ldots, \mathbf{x}^{(t+c)})$, where $(\mathbf{x}^{(t+1)}, \ldots, \mathbf{x}^{(t+c)})$ is an attractor, the part of the trajectory preceding the attractor, $(\mathbf{x}^{(1)}, \ldots, \mathbf{x}^{(t)})$ is called the \textit{transient}, with \textit{transient length} $t$.
All configurations that converge to the same attractor are said to form its \textit{basin of attraction}.

The \textit{transition graph} of the dynamical process $F$ on a given graph $G$ is the directed graph $\mathcal T =(V',E')$ where every configuration of the original graph is a node: $V' = S^n$.
There exists a directed edge $(\mathbf x^{(1)},\mathbf x^{(2)}) \subset E'$ if and only if the transition $F(\mathbf x^{(1)}) = \mathbf x^{(2)}$ exists.

\subsection{Analyzing the Dynamics}\label{sec:observables-notation}

The goal of this work is to analyze the properties of the dynamical process. Some of the questions we are interested in are: What are common properties of the basin of attraction of a given type of attractor? 
How are the limit cycles of the dynamical process composed, e.g. are they homogeneous and take the same value? 
What size do the limit cycles have? 

\paragraph{Dynamical transition graph motifs.} We want to answer these questions through the main object of this work's analysis, the \textit{backtracking attractor}, a dynamical motif of the update functions transition graph defined on the set of all possible configurations. 

Concretely, we define a $(p,c)$ backtracking attractor to be a trajectory 
\begin{align}
    \underline{\mathbf{x}} = (\mathbf{x}^{(1)},\ldots,\mathbf{x}^{(p)}, \mathbf{x}^{(p+1)},\ldots, \mathbf{x}^{(p+c)}) \in S^{n\times (p+c)}
\end{align}
with an incoming path of length $p$ ending in a limit cycle of length $c$, following the formalism introduced in~\cite{BDCM}.
This is a dynamical motif in the transition graph $\mathcal T$.
Note that we may also set $c=0$ to select paths without constraining them to a cycle. This is known as the (forward) dynamical cavity method \cite{Kanoria_2011-DCM}.
We may also set $p=0$ to individually analyze the limit cycles, as previously done for dense graphs by \cite{Hwang_2019}.

The properties of the process we are interested in are translated into the \textit{observables}, which are functions of backtracking attractors that return a scalar value.
We define them as an intensive quantity in the graph size which factorizes on the nodes, $\xi$,  or on the edges, $\gamma$, as
\begin{align}
   \xi & := \frac{1}{n} \sum_{i \in V} \Xi_i(\underline x_i) = \frac{1}{n}\Xi(\underline{\mathbf{x}})\,,\\\label{eq:observable}
   \gamma &:= \frac{1}{m} \sum_{a \in E} \Gamma_a(\underline{\mathbf x}_a) = \frac{1}{m} \Gamma(\underline{\mathbf{x}}) \,.
\end{align}
where all functions map to a single scalar value and may be specific to a given node or edge (via dependence on $i,a$).
Concretely, a node observable could measure the prevalence of a given state at initialization, or in the limit cycle.
An edge observable could quantify whether an edge contains nodes that are all in the same state or not.

In the following, the goal of our analysis is to quantify how many backtracking attractors are contained in the transition graph for a given graph $G$ and a dynamical process $F$.
Observables can be used to both condition the backtracking attractors that we take into account on a given value, and to measure a given observable for a selected subset of those backtracking attractors.
As an example, one could aim to find the average initialization (measured observable) of processes that end up in a homogeneous state (conditioned observable) after 5 steps (setting $p=5,c=0$).

\paragraph{Measuring backtracking attractors.}
In order to count the prevalence of the given backtracking attractors in a transition graph, we define a probability distribution over the dynamical motifs which only takes into account valid distributions and weighs them according to the observable of interest.

Concretely, for a given graph $G=(V,E)$ of size $n$, update rule $F$ and natural numbers $p,c$, we introduce a probability distribution over all trajectories of size $(p+c)$, $\underline{\mathbf{x}}\in S^{n\times (p+c)}$, with:
\begin{equation}
    P(\underline{\mathbf{x}})=\frac{1}{Z}\mathbbm{1}\left\{F(\mathbf{x}^{(p+c)})=\mathbf{x}^{(p+1)}\right\}\prod_{t=1}^{p+c-1}\mathbbm{1}\left\{F(\mathbf{x}^{(t)})=\mathbf{x}^{(t+1)}\right\}\,.
    \label{eq:prob-backtrack}
\end{equation}
Only a trajectory that is coherent with the update rule $F$ and ends in a limit cycle of length $c$ has a non-zero measure in this distribution -- precisely a $(p,c)$ backtracking attractor. 
The partition function $Z$ is introduced to normalize the distribution $P$. 
When $c=0$ we neglect the boundary condition, the first factor in \eqref{eq:prob-backtrack}, and give non-zero measure to all trajectories of length $p$. 
Finding the partition function, or equivalently the free entropy density $\Phi=\frac{1}{n}\log{Z}$, amounts to counting the number of valid $(p,c)$ backtracking attractors.

Beyond counting how many backtracking attractors exist for a given $p,c$ our analysis also allows to measure and condition on observables \eqref{eq:observable} of the attractors. To do so, we introduce them as flexible constraints, re-weightings of the probability distribution, in the form of Lagrange multipliers $\lambda$ in the distribution \eqref{eq:prob-backtrack}, with one multiplier per observable. For example, the  partition function of the modified distribution for a set of $l=1,\ldots,L$ observables $\Xi^{(l)}$ with intensive properties $\xi^{(l)}$ reads:
\begin{equation}\label{eq:partition-func}
    Z(\lambda_1,\ldots,\lambda_L)=\sum_{\underline{\mathbf{x}}}\exp\left(-\sum_{l=1}^L\lambda_l\Xi^{(l)}(\underline{\mathbf{x}})\right)\mathbbm{1}\left\{F(\mathbf{x}^{(p+c)})=\mathbf{x}^{(p+1)}\right\}\prod_{t=1}^{p+c-1}\mathbbm{1}\left\{F(\mathbf{x}^{(t)})=\mathbf{x}^{(t+1)}\right\}\,.
\end{equation}
We define the entropy $s$ of a class of backtracking attractors as $\mathcal{N}(\xi^{(1)},\ldots,\xi^{(L)})=e^{ns(\xi^{(1)},\ldots,\xi^{(L)})}$. The term $\mathcal{N}(\xi^{(1)},\ldots,\xi^{(L)})$ is the number of backtracking attractors in the transition graph.
As we take the graph size $n$ to infinity, we can apply the saddle point method to yield 
\begin{align}
    Z(\lambda_1,\ldots,\lambda_L)&=\int \left(\mathcal{N}(\xi^{(1)},\ldots,\xi^{(L)})e^{-n\sum_{l=1}^L\lambda_l\xi^{(l)}}\right)d\xi^{(1)}\ldots d\xi^{(L)}\\
    &\xrightarrow{n\rightarrow\infty} \max_{\xi^{(1)}, \ldots, \xi^{(L)}}e^{n\left(s(\xi^{(1)}, \ldots, \xi^{(L)})-\sum_{l=1}^L\lambda_l\xi_l\right)}\,,
\end{align}
which gives the free entropy density as
\begin{align}\label{eq:saddle-point}
\Phi(\lambda_1,\ldots,\lambda_L)=s(\hat{\xi}^{(1)},\ldots,\hat{\xi}^{(L)})-\sum_{l=1}^L\lambda_l\hat{\xi}^{(l)}\,,\\
\left.\frac{\partial s(\xi^{(1)},\ldots,\xi^{(L)})}{\partial \xi^{(l)}}\right|_{(\xi^{(1)},\ldots,\xi^{(L)})=(\hat{\xi}^{(1)},\ldots\hat{\xi}^{(L)})}= \lambda_l\,.
\end{align}
From \eqref{eq:saddle-point}, as $n\rightarrow\infty$ the observables $\Xi^{(l)}$ concentrate around:
\begin{equation}
     \frac{1}{n}\left\langle\Xi^{(l)}\right\rangle=\frac{1}{n}\sum_{\underline{\mathbf{x}}} \Xi^{(l)}(\underline{\mathbf{x}})\, P(\underline{\mathbf{x}})=-\frac{\partial{\Phi(\lambda_1,\ldots,\lambda_L)}}{\partial \lambda_l}=\hat{\xi}_l\,.
     \label{eq:obtain-observable}
\end{equation}
Here, $\langle \cdot \rangle$ denotes the average over the distribution $P$. If one can compute $\Phi(\lambda_1,\ldots,\lambda_L)$ and $\hat{\xi}^{(l)}=\frac{1}{n}\langle\Xi^{(l)}\rangle$ for a given set of $\lambda_l$, then $s(\hat{\xi}^{(1)},\ldots,\hat{\xi}^{(L)})$ obtained from \eqref{eq:saddle-point} gives the entropy of the class of backtracking attractors defined by the constraints $\frac{1}{n}\Xi^{(l)}(\underline{\mathbf{x}})=\hat{\xi}^{(l)}$, $l=1,\ldots,L$. The same holds for observables defined on the edges, if the edges scale as a constant fraction in the number of nodes.

By setting $\lambda_{l}=0$ it is possible to measure the value of the observable without constraining it. In other words, $\hat{\xi}^{(l)}$ is the value of the observable that is most likely to be observed in a typical trajectory of the dynamical process whose transient length is smaller or equal to $p$ and limit cycle length is a divisor of $c$. 

While these quantities are defined easily mathematically, computing the exponential sum in the partition function $Z$ in \eqref{eq:partition-func} is computationally infeasible. Hence, we will introduce the cavity method later on which approximates the free entropy density in a manner that is exact in the leading order for large random hypergraphs. Before this, we turn to the $k$-XOR-SAT problem and its interesting properties for motivation.

\section{The \texorpdfstring{$\boldsymbol{k}$}{k}-XOR-SAT Problem and Quench Dynamics}\label{sec:XOR-SAT}
The main focus of this work is a special subclass of dynamical processes on graphs. These processes on the graphs relate both to the $k$-XOR-SAT problem and the diluted (sparse) $p$-spin problem. The dynamics are such that each node synchronously chooses a state that satisfies the desired properties of the global problem -- but only based on its local neighborhood. Fixed points of the dynamical process are solutions to the problem.
Our goal is to determine when these fixed points are (not) reached and how long this typically takes when we initialize the dynamical process at random instances of the problem.

\subsection{The \texorpdfstring{$k$}{k}-XOR-SAT / diluted \texorpdfstring{$p$}{p}-spin Problem}
We consider the special case of hypergraphs where the nodes take two possible states $S=\{\pm 1\}$. Every node represents a boolean variable, where we take $+1$ to be \textit{true} and $-1$ to be \textit{false}. 
We interpret each hyperedge $a$ that connects a number of such nodes as a logical expression, a clause. For a given configuration $\mathbf{x}$, we attribute a truth value to each hyperedge, which is the structure equivalent to a clause in XOR-SAT. The goal is to have all clauses satisfied. 

\paragraph{Problem Statement.} We consider the case where a hypergraph edge $a=\{i_1,\ldots,i_{k_a}\} \subseteq V$ takes the parity of the incoming variables as
\begin{align}
    y_a := J_{a}\prod_{i\in a}x_i\,.
\end{align} 
Here, we can view $J_{a}\in\{\pm 1\}$ as the desired parity of the interaction, a property that is local to the hyperedge~$a$. 
To each configuration $\mathbf{x}$, we attribute an energy value which relates to the number of unsatisfied clauses using the following energy density:
\begin{equation}
    e(\mathbf{x}) := -\frac{1}{m}\sum_{a\in E} y_a\,.
    \label{eq:energy_def}
\end{equation}
which takes values in $[-1,+1]$.
This definition corresponds to setting a satisfied clauses value at $y_a=1$ and each unsatisfied clause at $y_a=-1$, and normalizing by the number of hyperedges $|E|=m$ in the graph. When we consider $d$-regular and $k$-uniform graphs we have $|E|=\frac{nd}{k}$.
It is $-1$ if all clauses are satisfied. 
When it is possible to find a configuration $\mathbf{x}$ which takes value $-1$, we say that the problem is satisfiable for a given graph $G$ and constraint set $\mathbf J 
\in \pm 1^n$.

Note that our formulation of the problem does not incorporate variables that appear negated in a clause, so this formulation does not incorporate every possible XOR-SAT problem. However, universality results along the lines of \cite{spinglass-antiferromagnetic} imply that the results we obtain also have close implications for the versions of the $k$-XOR-SAT problem where variables are randomly negated in clauses.  

\paragraph{Related Formulations.}
For $d$-regular and $k$-uniform graphs, this class of problems is known in statistical physics as the diluted $p$-spin model. Here, $p$ corresponds to $k$ in our notation, and diluted highlights that the analysis is for sparse graphs -- $d$ is constant in $n$.
More specifically, the cases of different distributions over $J_a$ are known as
\begin{table}[h]
    \centering
    \begin{tabular}{rl}
        \textit{spin-glass} & for $J_a = \pm 1$ uniformly at random, \\
        \textit{ferromagnet} & for $J_a = +1$ everywhere, \\
        \textit{anti-ferromagnet} & for $J_a = -1$ everywhere. 
    \end{tabular}
\end{table}

\paragraph{Static Properties.} For even values of $k$, the conjecture introduced in~\cite{spinglass-antiferromagnetic} suggests a universality between all three formulations. 
It was found that $k$-XOR-SAT instances are satisfiable with high probability when the ratio $\alpha=\frac{d}{k}$ is greater than $1$~\cite{PITTEL_SORKIN_2016}. 
When $k$ is odd, note that there is a trivial ground state (configuration that minimizes the energy density) for the ferromagnet with $J_a=1$ (anti-ferromagnet with $J_a=-1$) by setting all nodes to $+1$ ($-1$). 
In both cases, the non-trivial transition energy between the Replica Symmetric phase (RS) and the 1-step Replica Symmetry Breaking (1RSB) is in agreement with the result of the spin glass model. The former is shown in Appendix~\ref{app:RSB}, and the latter was computed in~\cite{k=3}. This allows us to further argue in favor of equivalence between the two models both for even and odd values of $k$, apart from the trivial $+1$ ($-1$) configuration. Instead of analyzing the static properties, in the following we turn to the analysis of a specific algorithm: The quench.

\subsection{Synchronous Quenched Local Updates}

We consider the \textit{quench dynamics} on $k$-XOR-SAT instances. The idea is that this process behaves as a local greedy algorithm seeking to minimize the energy of the system. A quench corresponds to a system that is initialized at a high temperature and abruptly cooled down to zero temperature. It then evolves in a greedy and synchronous way to locally minimize the energy, until it reaches a local minimum or limit cycle, hence the name for the dynamics we investigate.

\paragraph{The update function.} At initialization $t=1$, the configuration $\mathbf{x}^{(1)}$ of a given hypergraph is chosen with every node taking values $\pm1$ uniformly at random. Each node is then updated to locally maximize the number of satisfied clauses based on the current configuration. 
To do so, we set 
\begin{align}
    f_i^{\innerF}(x_i;\mathbf{x}_{a\backslash i}):=x_i y_a = J_a \prod_{j \in a \setminus i } x_j\,.
\end{align}
This inner local update rule returns the value of $x_i$ that would satisfy the clause $a$ given that all other nodes where kept constant.
Since every node appears in several clauses, we aggregate the preferences over every individual clause via a majority voting mechanism.
This aggregation mechanism is defined as 
\begin{align}
f_i^{\outerF}\left(x_i;\left\{f_i^{\innerF}\left(x_i;\mathbf{x}_{a\backslash i}\right)\right\}_{a\in\partial i}\right):=\sign\left(x_i+\sum_{a\in\partial i}f_i^{\innerF}(x_i;\mathbf{x}_{a\backslash i})\right)\,
\end{align} 
Adding the current value of $x_i$ in the sum allows to break ties in favor of the current value of the node.

\paragraph{Specific observables.}
In Section~\ref{sec:observables-notation} we described how observables of the system can help to characterize different dynamical processes. In the case of the quench dynamics on the $k$-XOR-SAT we ask whether we can reach a low-energy configuration, what the attractors of the system look like, and how this depends on the initialization.
Defining a $(p,c)$ backtracking attractor of the dynamical system as $\underline{\mathbf{x}}:=(\mathbf{x}^{(1)}, \ldots, \mathbf{x}^{(p)}, \mathbf{x}^{(p+1)}, \ldots, \mathbf{x}^{(p+c)})$ on a hypergraph $G=(V,E)$ with $n$ nodes and $m$ hyperedges, we investigate the following observables:
\begin{align*}
\begin{array}{r l l}
    m_{\initF}(\underline{\mathbf{x}})&:=\frac{1}{n}\sum_{i\in V}x_i^{(1)} &\textrm{magnetization of the initial configration};\\\\
    m_{\attr}(\underline{\mathbf{x}})&:=\frac{1}{c}\sum_{t=p+1}^{p+c}\frac{1}{n}\sum_{i\in V}x_i^{(t)} &\textrm{average magnetization of the attractor;}\\\\
    e^{(t)}(\underline{\mathbf{x}})&:=- \frac{1}{m}\sum_{a\in E}J_a\prod_{i\in a}x_i^{(t)}&\textrm{energy of the configuration at time t;}\\\\
    e_{\attr}(\underline{\mathbf{x}})&:=\frac{1}{c}\sum_{t=p+1}^{p+c}e^{(t)}(\underline{\mathbf{x}}) &\textrm{average energy of the attractor;}\\\\
    \rho(\underline{\mathbf{x}})&:=\frac{1}{n}\sum_{i\in V}\mathbbm{1}\left\{\left|\sum_{t=p+1}^{p+c}x_i^{(t)}\right|<c\right\} & \textrm{nodes that do not stay constant during the attractor.}
    \end{array}
\end{align*}
In the remainder of this work we refer to nodes that change during a limit cycle as \textit{rattlers}.

\section{Dynamical Cavity Method for Hypergraphs}\label{sec:BDCM}

This section introduces the backtracking dynamical cavity method for the analysis of dynamical processes on hypergraphs, as a tool to study the evolution and convergence behavior of the system. Using this method, we can derive closed-form equations that predict the observables for a given update function in a high-dimensional limit, where the number of nodes grows to infinity for constant $d$ and $k$.

In Section~\ref{sec:observables-notation} we showed that the entropy obtained from \eqref{eq:saddle-point} gives a measure of the number of trajectories that converge in $p$ or fewer steps to an attractor of given period $c$. In the limit $p\rightarrow\infty$, this set of trajectories represents the basin of attraction of length-$c$ limit cycles.
By measuring observables on this basin of attraction, we aim to extract information about the dynamics of the process and its attractors.
At the same time, by only considering paths with $c=0$, i.e. not constraining them to end in cycles, we can explore the start of the dynamics from a random typical configuration and measure observables.
However, computing both the entropy and the observables directly is infeasible since it involves high-dimensional sums over the space of all possible graph configurations. 

In this section, we show how to apply the\textit{ cavity method}~\cite{mezard2001bethe} to the distribution of $(p,c)$ backtracking attractors $P(\underline{\mathbf{x}})$ to approximate the free entropy density $\Phi(\lambda_1,\ldots,\lambda_L)$ on a hypergraph. 
The idea of this method is to factorize the probability distribution from Section~\ref{sec:observables-notation} on the underlying graph structure and to then compute an approximation of the complete distribution via local message passing. 

For $c=0$ this is known as the \textit{dynamical} cavity method (DCM) and was introduced in its forward version in~\cite{Neri_2009-DCM,Lokhov_2015-DCM,Mimura_2009-DCM,Kanoria_2011-DCM,Hatchett_2004-DCM}. The \textit{backtracking} dynamical cavity method (BDCM) was developed in~\cite{BDCM} to consider backtracking attractor with $c>0$, which was inspired by work aiming to find variable-length limit cycles~\cite{Hwang_2019}.
In all cases, these methods were defined for simple graphs -- in this work we extend them to the more general setting of hypergraphs.

\subsection{General Hypergraphs}
For the cavity method to be applicable, the distribution $P(\underline{\mathbf{x}})$ must factorize on a graph.
We assume that all observables of interest can be written as a sum of local variables depending on a single edge, or a node and its neighborhood, as shown in \eqref{eq:observable}. Then, assuming there is one node observable $\Xi(\underline{\mathbf{x}})$ and one hyperedge observable $\Gamma(\underline{\mathbf{x}})$, the factorized probability distribution from \eqref{eq:prob-backtrack} reads:
\begin{equation}\label{eq:prob}
    P(\underline{\mathbf{x}})=\frac{1}{Z}\prod_{i\in V}\mathcal{A}_i\left(\underline{x}_i;\{\underline{\mathbf{x}}_{a\backslash i}\}_{a\in \partial i}\right)\prod_{a\in E}\mathcal{B}_a(\underline{\mathbf{x}}_a)\,.
\end{equation}
Here, $\mathcal{A}_i$ regroups the constraints and observables factorized on the nodes and $\mathcal{B}_a$ those factorized on the hyperedges for two Lagrangian parameters $\lambda,\gamma$:
\begin{align}
    \mathcal{B}_a(\underline{\mathbf{x}}_a)=&\exp\left\{-\lambda\Xi_a\left(\underline{\mathbf{x}}_{a}\right)\right\}\label{eq:B}\,,\\
    \mathcal{A}_i\left(\underline{x}_i;\{\underline{\mathbf{x}}_{a\backslash i}\}\right)=&\exp\left\{-\gamma\Gamma_a\left(\underline{x}_i;\{\underline{\mathbf{x}}_{a\backslash i}\}\right)\right\}\\&\mathbbm{1}\Bigl\{f_i\left(x_i^{(p+c)};\{\mathbf{x}_{a\backslash i}^{(p+c)}\}\right)=x_i^{(p+1)}\Bigr\}\prod_{t=1}^{p+c-1}\mathbbm{1}\Bigl\{f_i\left(x_i^{(t)};\{\mathbf{x}_{a\backslash i}^{(t)}\}\right)=x_i^{(t+1)}\Bigr\}\,,\label{eq:A}
\end{align}
where $\{\underline{\mathbf{x}}_{a\backslash i}\}$ is a shorthand notation for $\{\underline{\mathbf{x}}_{a\backslash i}\}_{a\in \partial i}$.
The factorized probability distribution \eqref{eq:prob} can be represented on a factor graph, where the nodes are the variable nodes and the hyperedges are factor nodes with function $\mathcal{B}_a$. Additional factor nodes are introduced to represent the constraints and observables factorized on the nodes, with function $\mathcal{A}_i$. Note that it is possible to add more observables on the nodes or the hyperedges if needed.

\begin{figure}[H]
    \centering
    \includegraphics[width=1\linewidth]{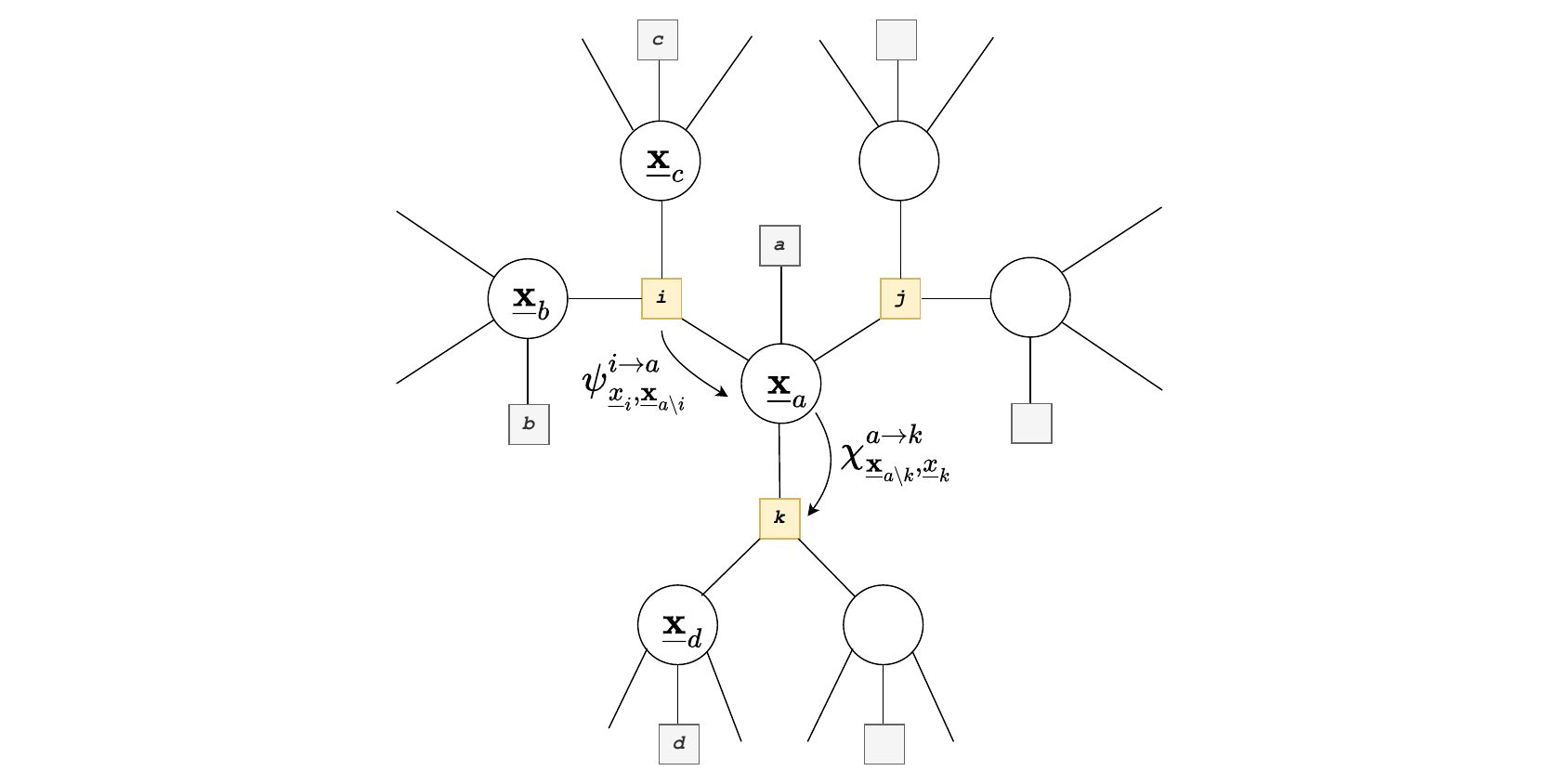}
    \caption{\textbf{Factor graph representation of the probability distribution \eqref{eq:prob} for 3-uniform 3-regular hypergraph}. The variable nodes can take value in $S^{k(p+c)}$. The factor nodes denoted by $i,j,k$ have a function $\mathcal{A}_i$ and enforce the evolution constraints defined by the local functions $f_i^{\innerF}$ and $f_i^{\outerF}$ and the chosen attractor constraints factorized on the nodes $\Xi_i$. The factor nodes denoted by $a,b,c$ have function $\mathcal{B}_a$ who implement the constraints factorized on the hyperedges $\Gamma_a$. The messages $\chi^\rightarrow$ and $\psi^\rightarrow$ are propagated in every direction between the variable and factor nodes of type $i,j,k$.
    }
    \label{fig:factor-graph-1}
\end{figure}

\paragraph*{Removing short loops.}If the underlying factor graph is a tree, applying the cavity method to this distribution allows one to exactly compute the properties of the distribution, such as its marginals and entropy. 
If the graph is similar to a tree, in the sense that it only contains large loops (logarithmic in the system size $n$), the cavity method has been shown to give accurate measurements.
However, in the representation selected before, any two nodes $i,j$ belonging to a same hyperedge $a$ are connected in the factor graph both through their respective factor nodes $\mathcal{A}_i\left(\underline{x}_i;\{\underline{\mathbf{x}}_{a\backslash i}\}_{a\in \partial i}\right)$ and $\mathcal{A}_j\left(\underline{x}_j;\{\underline{\mathbf{x}}_{a\backslash j}\}_{a\in \partial j}\right)$, and through the factor node $\mathcal{B}(\underline{\mathbf{x}}_a)$.
This results in short loops of length four.
This issue can be resolved by considering the hyperedges of the hypergraph as the variable nodes containing the values of all nodes belonging to the hyperedge. This modified factor graph architecture was similarly used in~\cite{Lokhov_2015-dual-factor-graph,Behrens_2022-dual-factor-graph,BDCM} to obtain a regular factor graph. 
Now, the nodes $i\in V$ are written as factor nodes with function $\mathcal{A}_i\left(\underline{x}_i;\{\underline{\mathbf{x}}_{a\backslash i}\}_{a\in \partial i}\right)$ and an additional factor node with function $\mathcal{B}(\underline{\mathbf{x}}_a)$ is added to each variable node $a\in E$. This factor graph is shown in Fig.~\ref{fig:factor-graph-1}.

\paragraph*{Message Passing.} The Belief Propagation (BP) equations corresponding to the probability distribution~\eqref{eq:prob} then read:
\begin{align}
    \chi^{a\rightarrow i}_{\underline{\mathbf{x}}_{a\setminus i},  \underline{x}_i}&=\frac{1}{Z^{a\rightarrow i}}\mathcal{B}_a(\underline{\mathbf{x}}_a)\prod_{j\in\partial a\backslash i}\psi^{j\rightarrow a}_{\underline{ x}_j,\underline{\mathbf{x}}_{a\setminus j}}\,,\\
    \psi^{i\rightarrow a}_{\underline{ x}_i,\underline{\mathbf{x}}_{a\setminus i}}&=\frac{1}{Z^{i\rightarrow a}}\sum_{\substack{\{\underline{\mathbf{x}}_{b\setminus i}\}\\b\in\partial i \backslash a}} \mathcal{A}_i\left(\underline{x}_i;\{\underline{\mathbf{x}}_{b\backslash i}\}_{b\in \partial i}\right)\prod_{b\in\partial i\backslash a}\chi^{b\rightarrow i}_{\underline{\mathbf{x}}_{b\setminus i},  \underline{ x}_i}\,.
\end{align}
Once they have converged, the BP messages $\chi^{\rightarrow}$ and $\psi^{\rightarrow}$ can be used to compute an approximation of free energy density under the factorized model. We call this Bethe free entropy density $\Phi_{\RS}$ to highlight that it was obtained under the replica symmetric assumption, i.e. under the assumption that the probability distributions over trajectories can be locally represented via belief propagation.
\begin{align}Z^a&=\sum_{\underline{\mathbf{x}}_a}\mathcal{B}_a(\underline{\mathbf{x}}_a)\prod_{i\in a}\psi^{i\rightarrow a}_{\underline{x}_i, \underline{\mathbf{x}}_{a \setminus i}}\,,\\
Z^i&=\sum_{\substack{\{\underline{\mathbf{x}}_a\}\\a\in\partial i}}\mathcal{A}_i\left(\underline{x}_i;\{\underline{\mathbf{x}}_{a\backslash i}\}_{a\in \partial i}\right)\prod_{a\in\partial i}\chi^{a\rightarrow i}_{\underline{\mathbf{x}}_{a\setminus i},\underline{x}_i}\,,\\
Z^{ai}&=\sum_{\underline{\mathbf{x}}_{a\setminus i}}\sum_{\underline{{x}}_i}\chi^{a\rightarrow i}_{\underline{\mathbf{x}}_{a\setminus i},\underline{x}_i}\,\psi^{i\rightarrow a}_{\underline{x}_i,\underline{\mathbf{x}}_{a\setminus i}}\,,\\
    N\Phi_{\RS}&=\log(Z)=\sum_{a\in E}\log(Z^a)+\sum_{i\in V}\log(Z^i)-\sum_{a\in E}\sum_{i\in a}\log(Z^{ai})\,.
\end{align}
Iterating the BP messages on a given graph, allows one to find a fixed point, for which the above Bethe free entropy can be evaluated.

\subsection{The case of \texorpdfstring{$d$}{d}-regular, \texorpdfstring{$k$}{k}-uniform hypergraphs}
In the case of a $d$-regular $k$-uniform hypergraphs with update functions $f^{\innerF}$ and $f^{\outerF}$, we assume that the probability distributions over the messages are independent of the concrete neighbourhood of node $i$.
Likewise, we assume that all observable functions $\Xi_i$ and $\Gamma_a$ are independent of the respective node or edge. Note that for the $k$-XOR-SAT problem this is the case only when we consider the (anti-)ferromagnetic case with $J_a$ constant in $a$, and \textit{not} the spin-glass.
Hence, we assume that all BP messages become the same and can be iterated on a single node. By loosing this dependence, the equations for the messages are simplified as
\newcommand{\bbX}{\underline{\mathbf{X}}}
\newcommand{\sX}{\underline{x}}
\begin{align}
    \chi^{\rightarrow}_{\bbX,\sX}&=\frac{1}{Z^\chi}\mathcal{B}\left([(\bbX)_1,\ldots,(\bbX)_{k-1},\sX]\right)\prod_{i = 1}^{k-1}\psi^{\rightarrow}_{[(\bbX)_1,\ldots,(\bbX)_{k-1},\sX]_{\setminus i}, (\underline{\mathbf X})_i}\label{eq:BP-chi-1}\,,\\
    \psi^{\rightarrow}_{\bbX,\sX}&=\frac{1}{Z^\psi}\sum_{[\tilde \bbX^{(1)}, \ldots, \tilde \bbX^{(d-1)}]}\mathcal{A}\left(\underline{x};[\tilde \bbX^{(1)}, \ldots, \tilde \bbX^{(d-1)},\bbX]\right)\prod_{a=1}^{d-1}\chi^{\rightarrow }_{\bbX^{(a)},\sX}\label{eq:BP-psi-1}\,,\\
\end{align}
where the $\bbX$ are defined to be $k-1$ tuples of size $S^{p+c}$ configurations, representing a neighborhood. Likewise $\sX$ is a single $S^{p+c}$ configuration. By the $\setminus i$ we mean that we take away the $i$
'th element of a tuple. These equations in turn lead to the Bethe entropy, where now $\bbX$ represents a $k$-tuple instead, as
\begin{align}Z^{\rm var}&=\sum_{\sX}\sum_{\bbX}\mathcal{B}\left([(\bbX)_1,\ldots,(\bbX)_{k-1},\sX]\right)\prod_{i = 1}^{k-1}\psi^{\rightarrow}_{[(\bbX)_1,\ldots,(\bbX)_{k-1},\sX]_{\setminus i}, (\underline{\mathbf X})_i}\,,\\
Z^{\rm fac}&=\sum_{\sX}\sum_{[\bbX^{(1)}, \ldots, \bbX^{(d)}]}\mathcal{A}\left(\underline{x};[ \bbX^{(1)}, \ldots, \bbX^{(d)}]\right)\prod_{a=1}^{d}\chi^{\rightarrow }_{\bbX^{(a)},\sX}\,,\\
    Z^{\rm norm}&=\sum_{\sX}\sum_{\bbX}\chi^{\rightarrow}_{\bbX,\sX}\cdot\psi^{\rightarrow}_{\bbX,\sX}\,,\\
    \Phi_{\RS}&=\frac{d}{k}\log{\left(Z^{\rm var}\right)}+\log{\left(Z^{\rm fac}\right)}-d\log{\left(Z^{\rm norm}\right)}\,.
\end{align}
The goal is now to find the fixed point of the messages again, which can be done by initializing randomly and then iterating the equations.
From \eqref{eq:BP-chi-1} and \eqref{eq:BP-psi-1}, we observe that the number of values that need to be iterated at each step is $2\cdot|S|^{k\cdot(p+c)}$. The complexity of iterating one value of $\psi^{\rightarrow}$ scales as $\Theta(\exp(k\cdot d))$. 
The complexity of this operation can be reduced by noting that the update rules are independent of the ordering of both the hyperedges in the neighborhood of a node and the other nodes in a hyperedge.
The sum in \eqref{eq:BP-psi-1} can therefore be computed using a dynamical programming algorithm~\cite{Torrisi_2022-dynmical-programming}. This reduces the complexity of iterating one value of $\psi^{\rightarrow}$ to $\Theta(k\cdot d)$. 
This leads to an algorithm for solving the equations iteratively, which takes $\Theta\left(kd\cdot|S|^{k\cdot(p+c)}\right)$. This is tractable for small values of $k$, $p$ and $c$.

\subsection{A computational simplification for \texorpdfstring{$k$}{k}-XOR-SAT}\label{sec:simplify-xor}

We want to apply this method for update functions which represent the $k$-XOR-SAT problem introduced in Section~\ref{sec:XOR-SAT}.
The specific problem allows for a further computational simplification: We can achieve an iteration cost of $\Theta((d+k)|S|^{p+c})$ which is no longer exponential in the hyperedge size $k$.

Note that we can reformulate the $k$-XOR-SAT problem in a way such that the hypergraph can be interpreted as an ordinary graph where hyperedges are a different type of nodes taking value $y_a=g(\mathbf{x}_a^{(t)})$ and edges connect the nodes to the hyperedges to which they belong. 

\begin{figure}
    \centering
    \includegraphics[width=1 \linewidth]{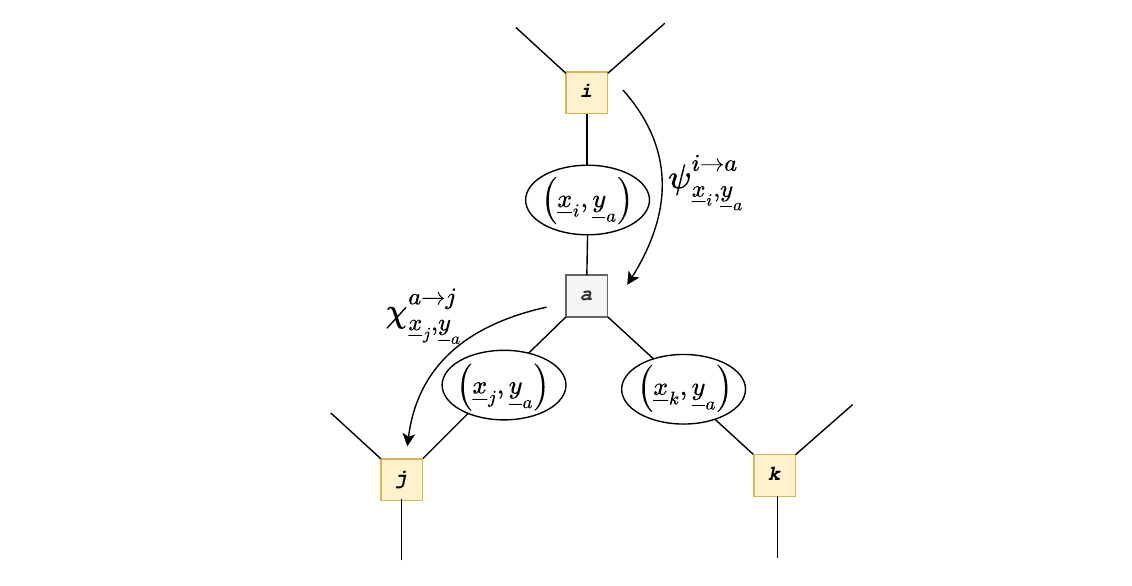}
    \caption{\textbf{Factor graph representation of the probability distribution \eqref {eq:prob-2} for a 3-uniform 3-regular hypergraph}. The variable nodes represent tuples $(i,a)$ of a node and a neighboring hyperedge, taking value in $S^{p+c}\times S^{p+c}$. The factor nodes denoted by $i,j,k$ have function $\mathcal{A}_i$ and the factor nodes denoted by $a,b,c$ have function $\mathcal{B}_a$. The messages $\chi^\rightarrow$ and $\psi^\rightarrow$ are propagated between the factor nodes.}
    \label{fig:factor-graph-2}
\end{figure}

The update of the graph is no longer done in a single step but takes place in two sequential steps (since the graph is bipartite over the two node types):
\begin{align}
    x_i^{(t+1)}&=f_i\left(x_i^{(t)}; \{y_a^{(t)}\}_{a \in \partial i}\right)=f_i^{\outerF}\left(x_i;\{f_i^{\innerF}(x_i^{(t)};y_a^{(t)})\}_{a\in\partial i}\right)\,,\\
    y_a^{(t+1)}&=g(\mathbf{x}_a^{(t+1)})\,,
\end{align}
where $g$ is a function that maps from a neighborhood of size $k$ to a value in $\pm 1$.
Notation-wise, note that we have $y_a \in \pm 1$,  $\mathbf{y} = \{\pm 1\}^m$ and $\mathbf{y}_i = \{y_a\}_{a \in \partial i}$. Again, an underline $\underline{y}_a$ considers the full trajectory over $p+c$ states.
Under this new formalism, \eqref{eq:prob}-\eqref{eq:A} become:
\begin{align}P(\underline{\mathbf{x}},\underline{\mathbf{y}})&=\frac{1}{Z}\prod_{i\in V}\mathcal{A}_i\left(\underline{x}_i;\underline{\mathbf{y}}_{i}\right)\prod_{a\in E}\mathcal{B}(\underline{y}_a; \underline{\mathbf{x}}_a)
    \label{eq:prob-2}\,,\\
    \mathcal{B}_a(\underline{y}_a; \underline{\mathbf{x}}_a)&=e^{-\lambda \Xi(\underline {\mathbf{x}}_{a})}\prod_{t=1}^{p+c}\mathbbm{1}\Bigl\{y_a^{(t)}=g(\mathbf{x}_a^{(t)})\Bigr\}\,,\\
    \mathcal{A}_i\left(\underline{x}_i;\underline{\mathbf{y}}_i\right)&=e^{-\gamma \Gamma(\underline{x}_i;\underline{\mathbf{y}}_i)}\mathbbm{1}\Bigl\{f_i\left(x_i^{(p+c)};\{y_a^{(p+c)}\}_{a\in\partial i}\right)=x_i^{(p+1)}\Bigr\}\prod_{t=1}^{p+c-1}\mathbbm{1}\Bigl\{f_i\left(x_i^{(t)};\{y_a^{(t)}\}_{a\in\partial i}\right)=x_i^{(t+1)}\Bigr\}\,.
\end{align}
Similarly to the previous section, the factor graph representation of the probability distribution contains loops of length 4, due to the fact that any couple $(i,a)$ of a node and a hyperedge in its neighborhood are connected through both their respective factor nodes $\mathcal{A}_i\left(\underline{x}_i;\underline{\mathbf{y}}_{i}\right)$ and $\mathcal{B}_a(\underline{y}_a; \underline{\mathbf{x}}_a)$. This issue is again solved by moving to the dual representation of the factor graph, taking the edges to be the variable nodes. In our representation, this results in the variable nodes being the tuples $(i,a)$ of a node and a hyperedge to which it belongs and each node and hyperedge being a factor node. The factor graph can be observed in Fig.~\ref{fig:factor-graph-2}.
BP equations can be written from this factor graph, using four different types of messages: two between variable nodes and factor nodes representing nodes $(i,a)\rightleftharpoons i$ and two between variable nodes and factor nodes representing hyperedges $(i,a)\rightleftharpoons a$. However, the messages $a\rightarrow (i,a)$ and $(i,a)\rightarrow i$, as well as $a\rightarrow (i,a)$ and $(i,a)\rightarrow i$, can be contracted to give the following compact BP equations:
\begin{align}
    \chi^{a\rightarrow i}_{\underline{x}_i,\underline{y}_a}&=\frac{1}{Z^{a\rightarrow i}}\sum_{\underline{\mathbf{x}}_{a\backslash i}}\mathcal{B}_a(\underline{y}_a; \underline{\mathbf{x}}_a)\prod_{j\in a\backslash i}\psi^{j\rightarrow a}_{\underline{x}_j,\underline{y}_a}\,,\\
    \psi^{i\rightarrow a}_{\underline{x}_i,\underline{y}_a}&=\frac{1}{Z^{i\rightarrow a}}\sum_{\underline{\mathbf{y}}_{i \setminus a}}\mathcal{A}_i\left(\underline{x}_i;\underline{\mathbf{y}}_{i}\right)\prod_{b\in\partial i\backslash a}\chi^{b\rightarrow i}_{\underline{x}_i,\underline{y}_b}\,.
\end{align}
The free entropy density can then be computed using the BP messages as:
\begin{align}
    Z^a&=\sum_{\underline{y}_a, \underline{\mathbf{x}}_a}\mathcal{B}(\underline{y}_a; \underline{\mathbf{x}}_a)\prod_{i\in a}\psi^{i\rightarrow a}_{\underline{x}_i,\underline{y}_a}\,,\\
    Z^i&=\sum_{\underline{x}_i,\underline{\mathbf y}_{i}}\mathcal{A}_i\left(\underline{x}_i;\underline{\mathbf{y}}_{i}\right)\prod_{a\in\partial i}\chi^{a\rightarrow i}_{\underline{x}_i,\underline{y}_a}\,,\\
    Z^{ai}&=\sum_{\underline{x}_i,\underline{y}_a}\chi^{a\rightarrow i}_{\underline{x}_i,\underline{y}_a}\psi^{i\rightarrow a}_{\underline{x}_i,\underline{y}_a}\,,\\
    n\Phi_{\RS}&=\sum_{a\in E}\log(Z^a)+\sum_{i\in V}\log(Z^i)-\sum_{a\in E}\sum_{i\in a}\log(Z^{ai})\,.
\end{align}
and the simplified equations in the case of a $d$-regular $k$-uniform hypergraphs with update functions $f^{\innerF}$ and $f^{\outerF}$ that are independent $i$ can be written as:
\newcommand{\bbY}{\underline{\mathbf{Y}}}
\begin{align}\label{eq:BP-chi-2}
\chi^{\rightarrow}_{\underline{x},\underline{y}}&=\frac{1}{Z^\chi}\sum_{\bbX}\mathcal{B}\left(\underline{y};[(\bbX)_1,\ldots,(\bbX)_{k-1},\underline{x}]\right)\prod_{i=1}^{k-1}\psi^{\rightarrow}_{(\bbX)_i,\underline{y}}\,,\\
\label{eq:BP-psi-2}
\psi^{\rightarrow}_{\underline{x},\underline{y}}&=\frac{1}{Z^\psi}\sum_{\bbY}\mathcal{A}\left(\underline{x};\left[(\bbY)_1,\ldots,(\bbY)_{d-1},\underline{y} \right]\right)\prod_{a=1}^{d-1}\chi^{\rightarrow}_{\underline{x},(\bbY)_a}\,,
\end{align}
where $\bbX$ is a $k-1$ tuple of trajectories $\underline y$ and $\bbY$ is a $d-1$ tuple of trajectories of $\underline x$. To write the entropy density, we slightly modify this notation to say that now $\bbX$ and $\bbY$ are $k$ and $d$ tuples of $\underline x$ and $\underline y$ respectively:
\begin{align}
Z^{\rm edge}&=\sum_{\underline{y}, \bbX}\mathcal{B}(\underline{y};\bbX)\prod_{i=1}^{k}\psi^{\rightarrow}_{(\bbX)_i,\underline{y}}\,,\\
Z^{\rm node}&=\sum_{\underline{x},\bbY}\mathcal{A}\left(\underline{x};\bbY\right)\prod_{a=1}^d\chi^{\rightarrow}_{\underline{x},(\bbY)_a}\,,\\
Z^{\rm norm}&=\sum_{\underline{x},\underline{y}}\chi^{\rightarrow}_{\underline{x},\underline{y}}\cdot\psi^{\rightarrow}_{\underline{x},\underline{y}}\,,\\
\Phi_{\RS}&=\frac{d}{k}\log{\left(Z^{\rm edge}\right)}+\log{\left(Z^{\rm node}\right)}-d\log{\left(Z^{\rm norm}\right)}\,.
\label{eq:free-entropy-2}
\end{align}
We conclude that in this setting the number of messages to be updated at each step has been reduced to $2\cdot|S|^{p+c}$ and, using a dynamical programming scheme to optimize the sums over $\bbY$ and $\bbX$ in \eqref{eq:BP-psi-2} and \eqref{eq:BP-chi-2}.
This amounts to $\Theta((d+k)|S|^{p+c})$ for a single iteration.
The exponential term in $k$ was reduced to a polynomial.

\section{Application to Quenches in \texorpdfstring{$k$}{k}-XOR-SAT}\label{sec:results}
In this section, we present results to characterize the evolution and convergence behavior of quench dynamics in $k$-XOR-SAT instances through the DCM and BDCM. We evaluate them based on their agreement with numerical simulations and compare them to the naïve mean-field approach, presented in Appendix~\ref{app:mean-field}. The code used to compute the BDCM and DCM results, the numerical simulations, and the mean-field predictions are available publicly at \url{https://github.com/SPOC-group/k-xorsat-bdcm}.

We have two principal questions that we want to answer for dynamics that are initialized at random: How fast do the dynamics reach a steady state or an attractor? What is the energy of the attractors they reach?

\subsection{General Phenomenology}

Depending on the degree and uniformity of the underlying graph, the behavior exhibited by the quench dynamics introduced in Section~\ref{sec:XOR-SAT} is quite different. From here on, we examine the anti-ferromagnetic case, where $J_a = -1$. For this, we can use the simplified and more efficient factor graph from Section~\ref{sec:simplify-xor}.
For an overview of the different dynamical processes, Table~\ref{tab:summary} groups different values of $d,k$ together when we find empirically that their behaviour is similar.
In particular, we distinguish \textit{even} and \textit{odd} degrees of $d$, and within the odd degrees we further distinguish between small and large hyperedge sizes $k$.

\begin{table}
    \centering
    \begin{tabular}{llp{2.5cm}p{5cm}p{5cm}}
        \toprule
        $d$ & $k$ & Transient length \par as $n\rightarrow\infty$ & Typical limit cycle as $n\rightarrow\infty$ & To be noted 
         \\
         \toprule\toprule even & any & $\Theta(\log n)$ & 1-cycles (larger cycles with vanishing fraction of rattlers as $n \to \infty$)  & behaviour of $k=2$ graphs generalizes to hypergraphs\\ \midrule
         odd & $2$  & $\Theta(\log n)$ & 2-cycles with average zero energy &  \\ \cmidrule{2-5}
         & $3$  & $\Theta(\log n)$ & large cycles that grow in $\Theta(n)$ & singular case with linear cycle length \\  \cmidrule{2-5}
          & $ \geq 4 $ & $\Theta(\exp n)$ & difficult to analyze empirically due to long convergence time \\ \toprule
    \end{tabular}
    \caption{\textbf{Overview: Quench dynamics of the XOR-SAT problem.} Convergence properties observed for $d$-regular $k$-uniform hypergraphs with different values of $d$ and $k$ and different tie-breaking rules. The notation $\theta\left(f(n)\right)$ is used to indicate that a quantity asymptotically behaves as $f(n)$, up to constants that can depend on other parameters of the problem such as $d$ and $k$. Evidence for this classification follows in this section and the appendix.}
    \label{tab:summary}
\end{table}

\subsection{Averaging the System: A Mean Field Approach}

One of the most common methods to analyse a dynamical process is a mean-field approach.
The key idea is to replace the interactions between individual elements of the system with the average interactions of the whole system.
The hope is that this ansatz holds true when the system is large enough.
In the case of sparse $k$-XOR-SAT, it is possible to apply this principle to the values $y_a$ defined on the hyperedges, and to update them in terms of the average probability of a given value for the whole graph, rather than in terms of the local interactions.
This leads to a prediction of the evolution of the energy of the process.
In Appendix~\ref{app:mean-field} we describe the simple iterative update equations that we derive from this principle in detail.

Fig.~\ref{fig:DCM} shows how the empirical evolution of 5 different samples of graphs evolves over time $t$, for different $d,k$.
We compare this with the prediction of the mean field approach, in solid lines.
During the first few timesteps the empirics and the mean field projections match closely. 
However, at some point the prediction diverges from the empirics as it reaches the homogeneous fixed point.
For many processes we examine here this mean-field fixed point is far off the actual fixed point that the process reaches empirically. This is not surprising as it does not incorporate enough knowledge of the correlation structure between neighboring nodes that builds up over the course of the process. 
In the following, we investigate the dynamical cavity methods and show that this method obtains more accurate predictions since it incorporates these correlations. We conjecture that in the large size limit of random hypergraphs, these methods capture all the correlations that influence leading order behaviour unless replica symmetry is broken, for which we have not observed any evidence in the studied problem. We examine both the forward analysis, as well as the backtracking one.

\subsection{The Beginning of the Process: Forward Dynamical Cavity Method}\label{sec:DCM}

\begin{figure}
    \centering
    \begin{subfigure}{.5\textwidth}
        \centering
        \includegraphics[width=220pt]{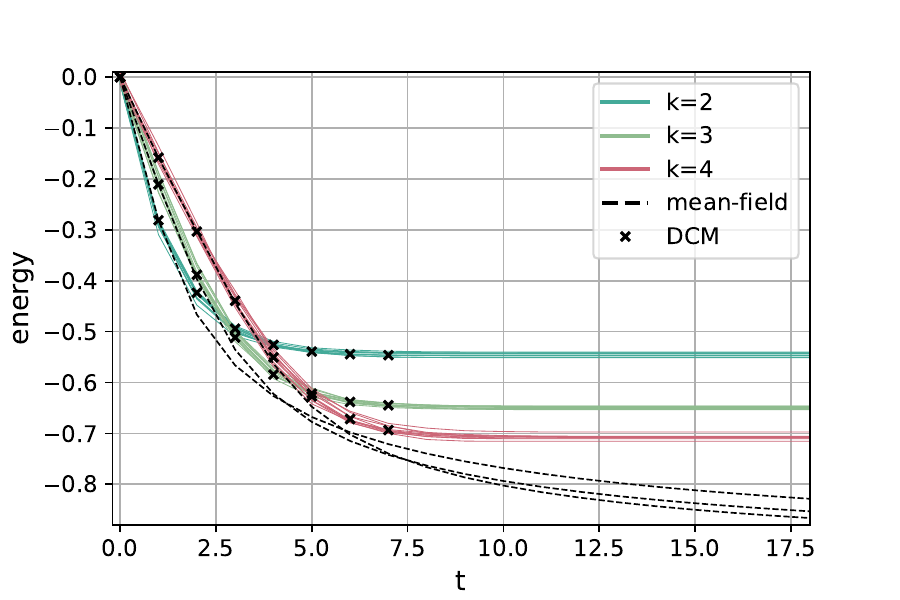}
        \caption{$d=4$}
        \label{fig:DCM-even}
    \end{subfigure}%
    \begin{subfigure}{.5\textwidth}
        \centering
        \includegraphics[width=220pt]{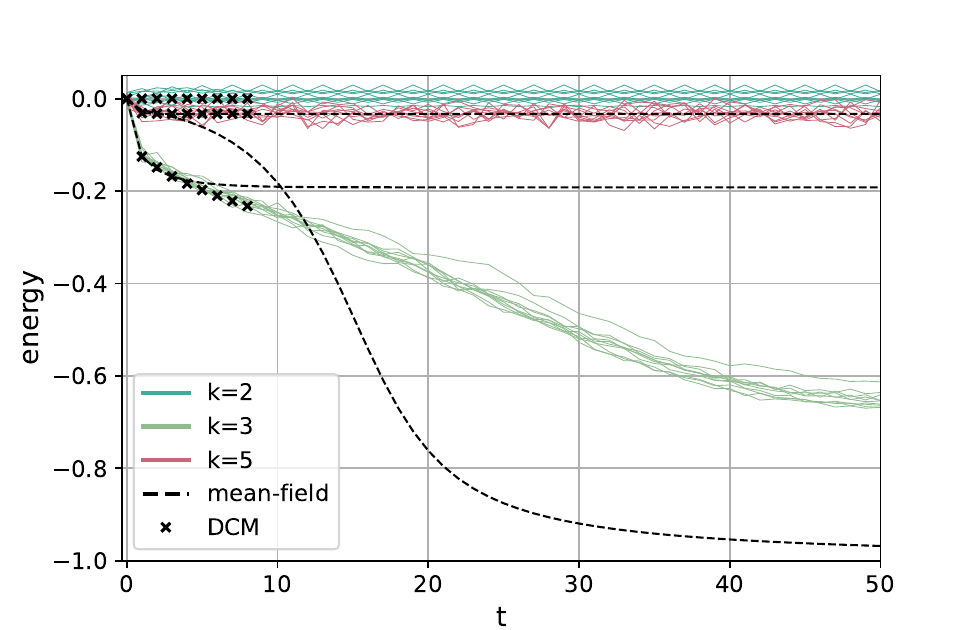}
        \caption{$d=3$}
        \label{fig:DCM-odd}
    \end{subfigure}
    \caption{\textbf{Evolution of the energy from numerical simulations compared to the dynamical cavity method and the mean-field model}. 10 numerical trajectories of the energy of the system simulated on random hypergraphs of size $n=10^4$ with degree (a) $d=4$, and (b) $d=3$ are shown in colors, for each value of the order $k$ considered. The simulations are initialized at random with zero magnetization. The energies obtained from the DCM with $p=7$ and the mean-field energy computed by iterating equations~\eqref{eq:mean-field-begin}-\eqref{eq:mean-field-end} are shown.}
    \label{fig:DCM}
\end{figure}

\paragraph*{The DCM is more accurate than the mean field.} Fig.~\ref{fig:DCM} shows the numerical evolution of the energy $e^{(t)}$ of the quench dynamics on hypergraphs of degree $d=3,4$ with hyperedges of varying size. We compare the mean-field prediction for the energy and the results of the DCM, computed from \eqref{eq:obtain-observable} after convergence of the BP messages for the first steps of the dynamics. The exponential dependency of the BP equations \eqref{eq:BP-chi-2}-\eqref{eq:BP-psi-2} on the length of the trajectory $p$ prevents us from obtaining numerical DCM results for $t\gtrsim 8$. Nonetheless, this suffices to show that the DCM accurately follows the numerical trajectories whereas the mean-field approximation already diverges.
With more computational power, this prediction could be continued over longer trajectories.

\paragraph{Even degree dynamics show fast convergence.}
In the case of a hypergraph with an even degree, here $d=4$, we observe in Fig.~\ref{fig:DCM-even} that the energy quickly decreases during the first steps of the process, before it plateaus and converges to a value that depends on the order $k$.  Qualitatively similar results are found for higher (even) degrees $d=6,8$.
The convergence time for hypergraphs with degree $d=4$ grows logarithmically with the graph size $n$, as numerically shown in Appendix~\ref{app:transient}. For the mean-field approximation, however, the energy trajectory does not stop decreasing and approaches the stable fixed point $-1$ as $t\rightarrow\infty$.

\paragraph*{Mean field is accurate for odd degree dynamics with chaotic and slow convergence.}
In the case $d=3$, i.e. hypergraphs with an odd degree, only the settings $k=2,3$ give dynamical processes whose transient lengths are logarithmic in $n$, as shown in Appendix~\ref{app:transient}. 
For any other $k>3$, the mean-field equations have a stable fixed point close to zero (between $-0.1$ and $0$), and the numerical trajectories are observed to chaotically oscillate around the mean-field fixed point, see Fig.~\ref{app::fig:mean-field-odd-4} in Appendix~\ref{app:additional-odd}. 
Their transient lengths grow exponentially with $n$, as shown in Fig.~\ref{app::fig:transient-odd} in Appendix~\ref{app:additional-odd}. 
Under those parameters, the dynamics are ineffective in reaching low-energy configurations. They adopt a chaotic behavior for an exponential time until they randomly visit a configuration belonging to an attractor. The qualitative behavior of their transient is correctly described by the mean-field equations, as can be seen in Fig.~\ref{fig:DCM-odd} and Fig.~\ref{app::fig:mean-field-odd-4}. However, Fig.~\ref{app::fig:mean-field-odd-4} shows that the mean-field and DCM energies for odd values of $k$ approach two distinct, although very close, fixed points as $p$ grows.
Because of the large transient length observed in this setting, the convergence of typical trajectories with $k>3$ cannot be studied either numerically or using the BDCM.

\paragraph*{Special cases for odd degrees.}
When $k=3$, the trajectories do not stay stuck near the mean-field stable fixed point $e\approx -0.192$. Unlike $k>3$, the energy keeps decreasing at a linear rate and plateaus slightly under $e=-0.6$. In Fig.~\ref{fig:DCM-odd}, one observes that the DCM correctly ignores the mean field plateau. On the other hand, for $k=2$, the dynamics get quickly trapped in cycles of length $2$ with energy close to zero. It was formally proven in \cite{proof-2-cycles} that quench dynamics on ordinary graphs, i.e. $k=2$, always converge to limit cycles of length at most $2$. On the other hand, large cycles ($c>100$) are numerically observed for $k=3,\ldots,7$. Given this qualitative difference, it would not be surprising to observe differences between the density of attractors around zero energy for $k=2$ and $k>2$. This might explain why trajectories on $2$-uniform hypergraphs quickly converge to limit cycles, whereas in the case $k>2$ the trajectories that get stuck at high energy need exponential time to find an attractor.

\subsection{Attractors of the Process: Backtracking Dynamical Cavity Method}

\begin{figure}
    \centering
    \begin{subfigure}{.5\textwidth}
        \centering
        \includegraphics[width=220pt]{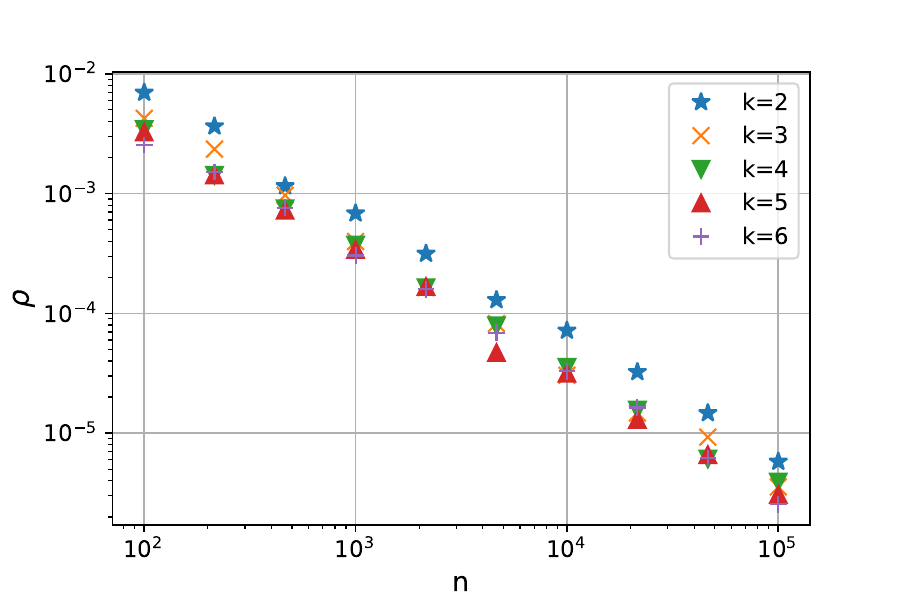}
        \caption{}
        \label{fig:1-cycles-numerics}
    \end{subfigure}%
    \begin{subfigure}{.5\textwidth}
        \centering
        \includegraphics[width=220pt]{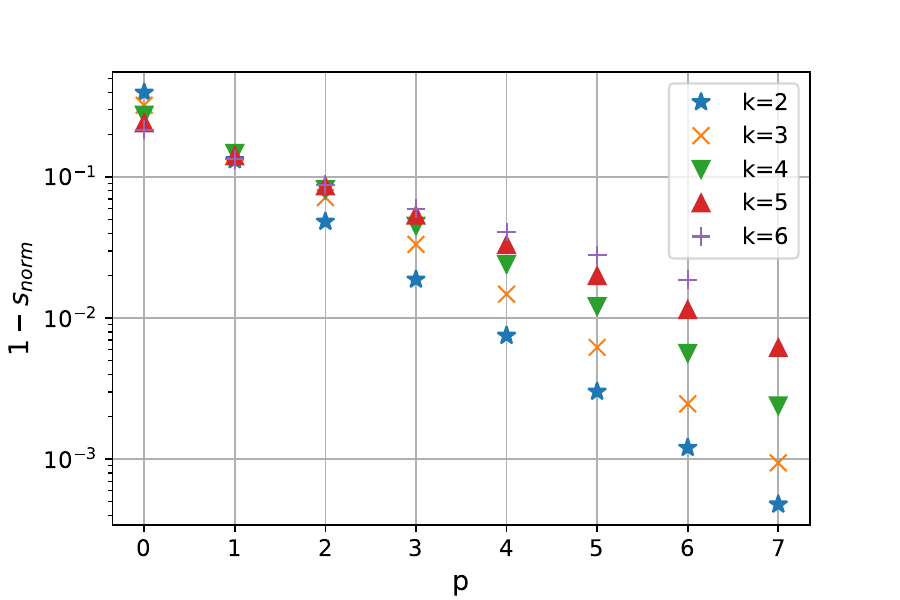}
        \caption{}
        \label{fig:1-cycles-BDCM}
    \end{subfigure}
    \caption{\textbf{Convergence to fixed points of trajectories on $4$-regular $k$-uniform hypergraphs}.
    (a) Numerical results - Fraction of rattlers, i.e. nodes that do not stay constant during a cycle, $\rho$, after convergence to an attractor, averaged over $7000$ simulations initialized at random with zero magnetization, as a function of the graph size $n$, (b) BDCM results - $1 - s_{\rm norm}$ with $s_{\rm norm}=\frac{s}{\log{2}}$ the normalized entropy of attractors of length $c=1$ as a function of the incoming path length $p$, obtained from \eqref{eq:free-entropy-2}.}
    \label{fig:even-d-1-cycles}
\end{figure}

Recall that the BDCM allows us to count the attractors of size $c$ of the dynamics with an incoming path length $p$.
We can use this property to obtain information on the basin of attraction of a specific type of attractor. 
Here we focus on fixed points, as they represent stable solutions.

As we increase $p$ in our analysis, the entropy of the number of backtracking attractors that we find grows: We capture a larger fraction of the basin of attraction -- i.e. the configurations that lead to these attractors within at most $p$ steps.
Recall that we want to answer what energetic states the dynamics reaches.
To do so,  we compute observables to obtain properties of these attractors that are reached from the basin of attraction -- here, we compute the energy of the fixed points.

Computing this value for $p \to \infty$ would encompass the full basin of attraction. However, this is computationally unfeasible. 
Therefore we compute the observables for small $p$ and with some care we extrapolate to larger $p$.
Because of the limitations of analyzing only short trajectories, we focus only on processes that take $\Theta(\log n)$ to converge.

\paragraph{Energy in the attractor for even degree.}
\begin{figure}[ht]
\centering
    \begin{subfigure}{.5\textwidth}
        \centering
        \includegraphics[width=220pt]{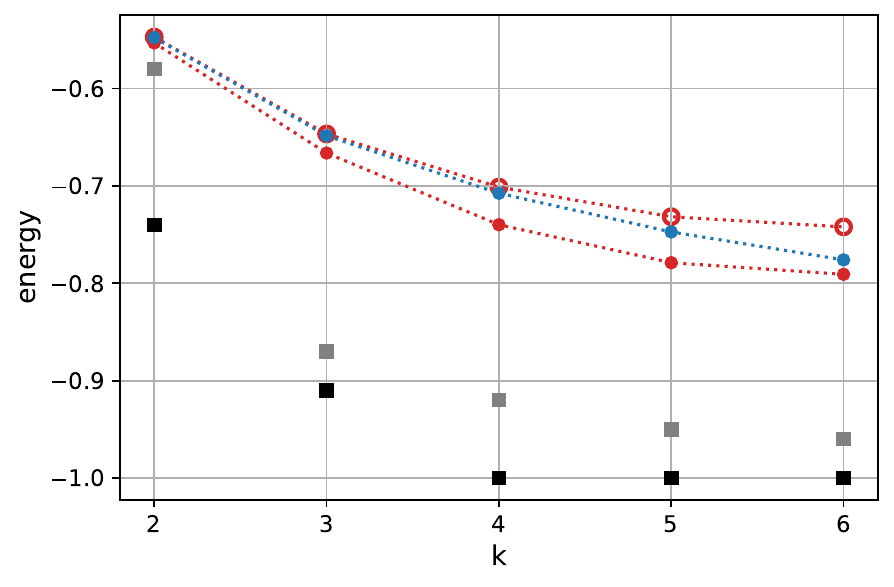}
        \caption{$d=4$}
        \label{fig:varying-k}
    \end{subfigure}%
    \begin{subfigure}{.5\textwidth}
        \centering
        \includegraphics[width=220pt]{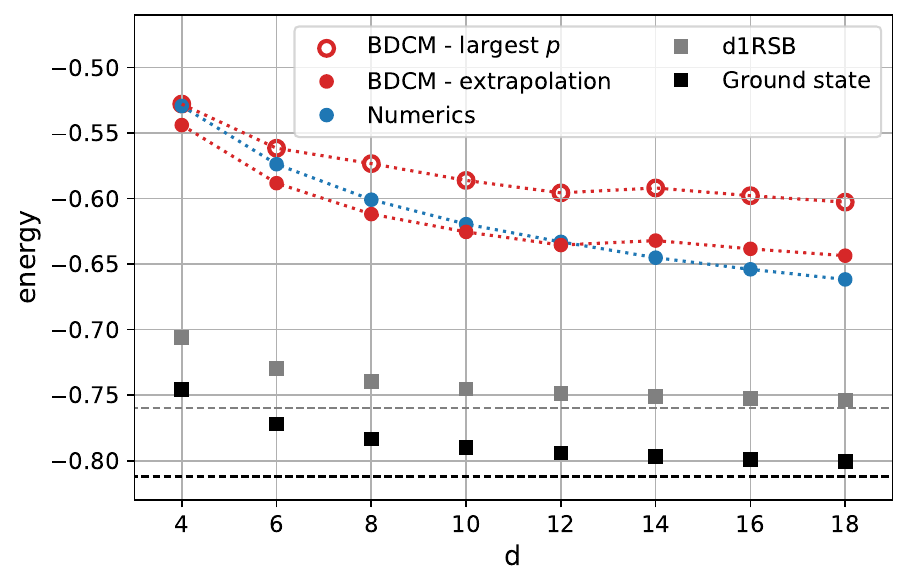}
        \caption{$k=3$}
        \label{fig:varying-d}
    \end{subfigure}
    
        \caption{\textbf{Energy of the attractor for typical trajectories obtained from the BDCM and numerical simulations}. Comparison of the BDCM result for the energy $e_{\attr}$ of a typical attractor of length $c=1$, with the energy obtained from numerical simulations initialized at random with zero magnetization performed for $n$ going from $10^2$ to $10^4$ and linearly extrapolated at $\frac{1}{n}\rightarrow 0$. For the BDCM, both the result obtained with the largest $p$ considered and the extrapolated value at $\frac{1}{p}\rightarrow 0$ are presented. The ground state energies and the energies at which the phase transition between replica symmetry and dynamical 1-step replica symmetry breaking occurs (d1RSB) for the $k$-XOR-SAT model with spin glass interactions are shown as a reference.
        (a)~The results are shown for $4$-regular hypergraphs with varying degree $k$.
        The procedure to compute the d1RSB energies is detailed in Appendix~\ref{app:RSB} and the ground state energies can be found in~\cite{spinglass-antiferromagnetic,k=3,PITTEL_SORKIN_2016}.
        (b)~The degree is fixed at $k=3$ and the results are plotted for growing degrees $d$. The energy is multiplied by $\sqrt{\frac{d}{2k}}$ compared to its definition in (\ref{eq:energy_def}) to make the limit $d\rightarrow\infty$ converge to the fully connected $p$-spin model~\cite{Wang_2021-fully-connected,Guiselin_2022-fully-connected,Montanari_2003-energies-k=3}. The ground state and 1dRSB energy for the fully connected model are shown as dashed lines. Those energies, both for growing $d$ and for the fully connected case, were computed in~\cite{Montanari_2003-energies-k=3} and are here multiplied by $\sqrt{\frac{k}{2d}}$ to match the rescaling of the BDCM and numerical energies.
    The complete results of the BDCM and the numerical simulations, that were used to obtain the extrapolated values, are shown in Fig.~\ref{app::fig:convergence-energy} in Appendix~\ref{app:additional-even}, and all BDCM results are detailed in Table.~\ref{app::tab:BDCM-even-degree-evenk-stay}. The uncertainty on the linear fits used to obtain the $\frac{1}{n}\rightarrow 0$ extrapolation for the numerical results are too small to be visually observed. The BDCM energy with the largest $p$ considered represents an upper bound for the exact value at $p\rightarrow \infty$.}
    \label{fig:BDCM-energies}
\end{figure}
Fig.~\ref{fig:even-d-1-cycles} shows evidence that the quench dynamics of $k$-XOR-SAT instances with degree $d=4$ almost always end up in cycles of length $1$, in the large $n$ limit. In Fig.~\ref{fig:1-cycles-numerics}, the fraction of rattlers $\rho$ of numerically observed attractors decreases as $\frac{1}{n}$  for $k=2,\ldots,6$. This empirical result aligns with the theoretical prediction in Fig.~\ref{fig:1-cycles-BDCM}. Using \eqref{eq:saddle-point} and the BDCM equations without Lagrange constraints ($\lambda=0$ and $\gamma=0$), it shows that the normalized entropy of attractors with length $c=1$ approaches $1$ exponentially as the incoming path length $p$ increases. The BDCM predictions for attractors of length $c=2$ are exactly identical to those of length $c=1$, indicating that in the large $n$ limit no attractor of length $c=2$ is reached from a non-negligible fraction of the initial configuration space. This result means that computing the energy of the attractor for the setting $c=1$ and growing values of $p$ gives the typical energy reached by a trajectory of this process on a $4$-regular hypergraph. We approximate the energy in the limit $p\rightarrow\infty$ by means of a linear fit of the values obtained from \eqref{eq:obtain-observable} with $\Xi(\underline{\mathbf{x}})=e_{\attr}(\underline{\mathbf{x}})$ and evaluated at $\lambda=0$ plotted as a function of $\frac{1}{p}$. The value of the fit evaluated at $\frac{1}{p}=0$ then gives an approximation to the value of $e$ when $p$ is very large. Because of the computational complexity of the BDCM algorithm with large $p$ values, the approximation was computed from the $p\leq 7$ results only. Its extrapolation over $p$ in Fig.~\ref{fig:BDCM-energies} must therefore be taken with caution. As the connectivity of the hypergraph increases, only smaller values of $p$ are computationally tractable which makes the results for large $k$ or $d$ less precise. Fig.~\ref{fig:BDCM-energies} shows that the extrapolated BDCM result for the energy is in reasonable agreement with the numerical simulations, extrapolated as $n\rightarrow\infty$. 

In Fig.~\ref{fig:varying-k}, we observe that the energy reached by the quench dynamics is very close to the 1dRSB phase transition for $k=2$, whereas for $k=3$ they differ by $\sim0.2$. The 1dRSB phase transition corresponds to the energy under which the equilibrium properties of the system are described with replica symmetry breaking~\cite{1RSB-first,1RSB-explanation,1RSB-explanation-2,koller2024counting}. 
Fig.~\ref{fig:varying-d} shows the energies for $k=3$ and increasing $d$, rescaled to correspond to the fully connected $p$-spin model~\cite{Wang_2021-fully-connected,Guiselin_2022-fully-connected,Montanari_2003-energies-k=3} in the limit $d\rightarrow\infty$. The results suggest that the energy of the quench stays stuck well above the 1dRSB energy even in the large $d$ limit. This might indicate the emergence of a large number suboptimal frozen structures during the dynamical process in the case $k=3$. For instance, if a well-connected subset of the hypergraph becomes static, i.e. enough hyperedges are satisfied for all node values belonging to them to be fixed, but a few hyperedges immersed in this frozen subset are unsatisfied, those unsatisfied hyperedges will be trapped until the end of the process. The number of such structures that can be found in random $3$-XOR-SAT instances was studied in~\cite{cocco2003approximate}. It was found that $\Theta(n)$ suboptimal structures exist with high probability in any instance of the XOR-SAT problem with $k=3$.

\paragraph{Energy for odd degree -- ($k=2$).}
We saw in Section~\ref{sec:DCM} that hypergraphs with $d=3$ and $k=2$ numerically converge to cycles with close to zero energy, i.e. they essentially do not move away from the average energy of a random initial configuration. 
Recall that hypergraphs with $k=2$ are simple graphs as considered previously in \cite{BDCM}.
In agreement with previous results, we recover that the normalized entropy associated with $c=2$ converges to $1$ as $p\rightarrow\infty$, whereas the $c=1$ result stays significantly lower than one, indicating that the dynamical process would converge almost always to $2$-cycles. Consistent with the numerical observations, the energies obtained for $c=2$ are zero for all considered $p$. Those results can be observed in Appendix~\ref{app:additional-odd}. It allows us to conclude that the quench dynamics in the system limit with parameters $d=3$ and $k=2$ almost always end in cycles of length $2$ and zero energy.

\begin{figure}
    \centering
    \begin{subfigure}{.5\textwidth}
        \centering
        \includegraphics[width=220pt]{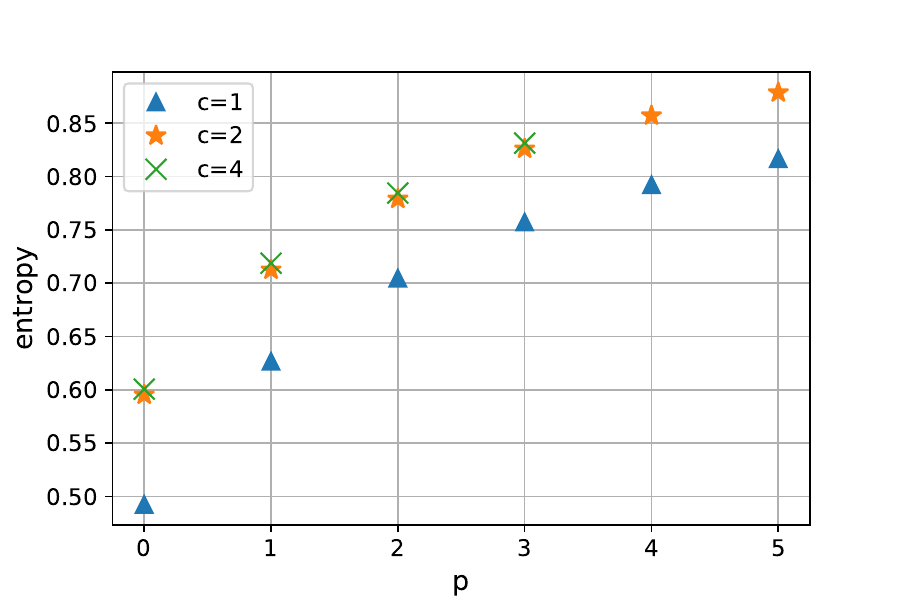}
        \caption{}
        \label{fig:entropy-k3}
    \end{subfigure}%
    \begin{subfigure}{.5\textwidth}
        \centering
        \includegraphics[width=220pt]{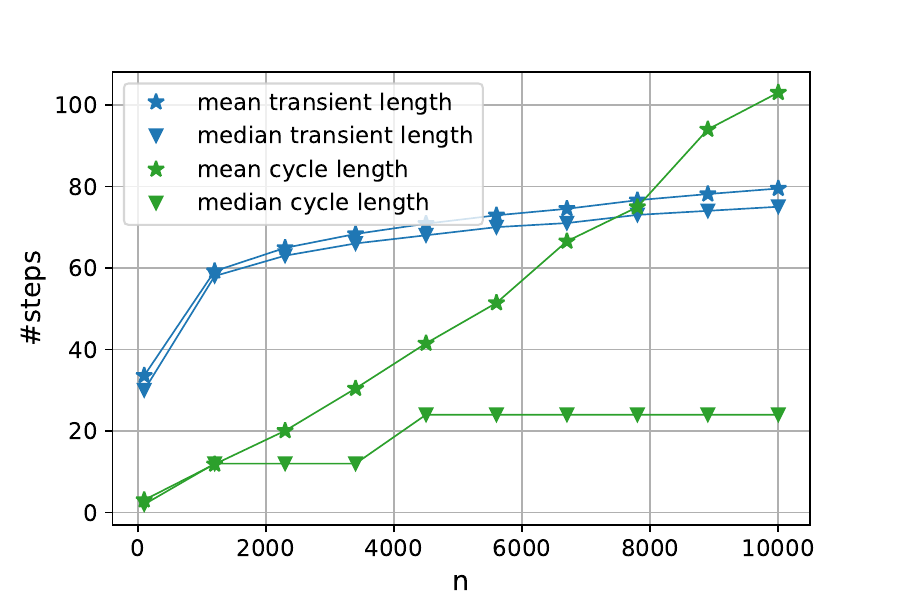}
        \caption{}
        \label{fig:k3-numerics}
    \end{subfigure}
    \caption{\textbf{Cycle and transient lengths of trajectories on $3$-regular $3$-uniform hypergraphs.} (a) BDCM results - Normalized entropy $\frac{s}{\log 2}$ of attractors of lengths $c=1,2,4$ as a function of the incoming path length $p$, obtained from \eqref{eq:free-entropy-2}. (b) Numerical results - Cycle lengths and transient lengths observed numerically as a function of the hypergraph size $n$. The mean and median are taken over $7000$ simulations initialized at random with zero magnetization.}
\end{figure}

\paragraph{Energy for odd degree -- ($k=3$).}
The case $k=3$ appears to be the only setting enabling hypergraphs of degree $d=3$ to reach low energy states with quench dynamics. However, one observes in Fig.~\ref{fig:entropy-k3} that the normalized entropy for small cycles $c=1,2,4$ does not quickly approach $1$ for growing values of the incoming path length $p$. Furthermore, the entropy of each considered $c$ is strictly greater than the smaller lengths. Those observations indicate that, unlike the results seen for hypergraphs with even degrees, the trajectories do not end in small attractors $c\leq 4$. The BDCM does not allow the study of the basin of attraction of larger attractors, as evaluating the equations numerically becomes exponentially costly in $c$. Fig.~\ref{fig:k3-numerics} shows the length of the cycles to which trajectories were numerically observed to converge and compares it to the transient length. Interestingly, for large enough $n$, all observed cycle lengths were found to be multiples of $12$, as presented in Fig.~\ref{app::fig:12-multiples} in Appendix~\ref{app:additional-odd}. In Fig.~\ref{fig:k3-numerics} we see that, while both the mean and the median of the transient lengths grow logarithmically in the considered range of hypergraph size $n$, the mean cycle length distances itself quickly from the median and becomes larger than the transient length. This suggests that the distribution of cycle lengths at large $n$ might be heavy-tailed.

\section{Conclusion}
In this work, we studied the quench dynamics of the XOR-SAT problem on $d$-regular $k$-uniform hypergraphs. A generalization of the BDCM introduced in \cite{BDCM} was developed to allow the study of dynamical processes on hypergraphs. Two versions of this extension were presented: a general version covering all synchronous, deterministic and locally defined update rules on hypergraphs, and a more specific one whose complexity depends linearly on the hyperedges size $k$, instead of exponentially.

In our results we showed that the DCM is more effective than a naive mean field approach which failed to predict update rules with logarithmic convergence time, whereas the DCM successfully predicted the initial phase of the quench trajectories. The BDCM was used to determine the type of attractor to which typical trajectories converge, for different values of the degree $d$ and the order $k$ of the hypergraphs. In the case $d=4$, it was found that almost all trajectories converge to fixed points. The BDCM enabled the evaluation of the performance of the quench dynamics to reach low-energy configurations. For the $3$-XOR-SAT, it was observed that, as $d$ grows, the quench stays stuck above the energy at which the transition from RS to 1dRSB occurs. In the specific case $d=3,\,k=3$, the dynamics were shown to converge in logarithmic time to large attractors, with plausibly a heavy-tailed distribution.

The generalization of the BDCM to hypergraphs opens many applications for studying dynamical properties of constraint satisfaction problems with interactions involving any number of variables. Examples of such problems involve the random K-SAT problem \cite{mezard2002analytic}, the random bicoloring aka NAE-SAT problem \cite{coja2012condensation}, or the occupation problems studied in \cite{zdeborova2008locked,zdeborova2008constraint}. This method provides a new perspective on the types of questions that can be addressed using the cavity method. However, further work would be needed to make the method applicable to stochastic processes, limiting the current applications to deterministic algorithms. The main limitation of the BDCM is its complexity growing exponentially with the length of the considered path.

\section*{Acknowledgments}
This research was supported by the NCCR MARVEL, a National Center of Competence in Research, funded by the Swiss National Science Foundation (grant number 205602).

\newpage
\begin{appendices}
\section{Mean Field Model}\label{app:mean-field}
We introduce a mean field approximation as a competing method to the DCM for predicting the energy evolution at the start of the process.
Starting from a random configuration of the system, where the node values are drawn independently from a Bernoulli distribution, the goal is to express the fraction of satisfied hyperedges at time $t$ in a recursive way. This enables us to iteratively compute the evolution of the energy of the system. The mean field approximation
neglects the local correlations between neighboring hyperedges and thereby allows one to write the evolution of the hyperedge values as a function of the fraction of satisfied constraints only.

We start by writing a discrete Master equation for the probability of a hyperedge $a$ to be satisfied at time $t+1$, i.e. $y^{(t+1)}=+1$.
The difference in the probability of a hyperedge $a\in E$ to be satisfied at time $t$ compared to the same probability at time $t+1$ can be decomposed: Into the probability that $y_a^{(t)}=-1$ ($a$ was unsatisfied) and then subsequently changed, minus the probability $y_a^{(t)}=+1$ ($a$ was satisfied) and subsequently changed. As a change in the value of the hyperedge at time $t$ is caused by an odd number of changes in the values of the nodes it contains at that very time, one can write:
\begin{align}\label{eq:update-before-mean-field}
    \mathbbm{P}\left(y_a^{(t+1)}=1\right) - \mathbbm{P}\left(y_a^{(t)}=1\right) =& \mathbbm{P}\left(\left.\sum_{i\in a}\mathbbm{1}\{x_i^{(t+1)}\neq x_i^{(t)}\}\;\mathrm{is\;odd\;}\right|y_a^{(t)}=-1\right)\mathbbm{P}\left(y_a^{(t)}=-1\right)\\
    &-\mathbbm{P}\left(\left.\sum_{i\in a}\mathbbm{1}\{x_i^{(t+1)}\neq x_i^{(t)}\}\;\mathrm{is\;odd\;}\right|y_a^{(t)}=1\right)\mathbbm{P}\left(y_a^{(t)}=1\right).
\end{align}
Following the mean-field approach, we assume the neighborhood of the nodes belonging to a hyperedge $a$ to be independent given the value of $y_a$, i.e. all edges $\left\{y_b\;|\;b\in \partial i\backslash a\right\}$.
Using this assumption, the probability of the number of nodes $i\in a$ changing value between time $t$ and $t+1$ being odd can be expressed as:
\begin{equation}
    \mathbbm{P}\left(\left.\sum_{i\in a}\mathbbm{1}\{x_i^{(t+1)}\neq x_i^{(t)}\}\;\mathrm{is\;odd\;}\right|y_a^{(t)}\right)=\sum_{\mathrm{odd }\;l\leq k} \mathbbm{P}\left(X_a^{(t)}=2l+1\right)\,.
\end{equation}
Here, $k$ is the number of nodes belonging to the hyperedge $a$ and $X_a^{(t)}$ is a binomial random variable.
It models a process with $k$ trials and a probability of success given by the probability of a node $i$ to change value between $t$ and $t+1$ given $y_a^{(t)}=1$, which we assume to be independent as
\begin{equation}
    \mathbbm{P}\left(\left.x_i^{(t+1)}\neq x_i^{(t)}\right|y_a^{(t)}\right)=\mathbbm{P}\left(\left.y_a^{(t)}+\sum_{b\in \partial i \backslash a}y_b^{(t)}<0\right|y_a^{(t)}\right)\,,
    \label{eq:probability-success}
\end{equation}
for any $i\in a$.
$X_a^{(t)}$ represents the number of nodes $i$ in $a$ that change value between $t$ and $t+1$ given $y_a^{(t)}$.

This formulation enables us to model the probabilistic evolution of the hyperedge values $y_a^{(t+1)}$ as a function of the previous values $y_a^{(t)}$ only, getting rid of the individual node values $x_i^{(t)}$.
One can then consider the dual of the hypergraph where the hyperedge values become the nodes and the neighborhood of the nodes represent links between several hyperedges.

The key idea of the mean field approximation is to replace the interactions with individual elements of the system by the average of the interactions with the whole system.
Applying this principle to the system of $y_a$ values, we replace the probability of a neighboring hyperedge $b$ to have a value $y_b$ by the fraction of hyperedges with value $y_b$ in the system $p^{(t)}=\frac{k}{nd}\sum_{a\in E}\mathbbm{1}{\{y_a^{(t)}=1\}}$, neglecting the dependency between the values of two neighboring edges.
The $y_b^{(t)}$ for $b\in \partial i \backslash a$ then become independent and identically distributed Bernouilli variables with probability $p^{(t)}$.
The probability of a node to flip at a given time step from \eqref{eq:probability-success} then becomes:
\begin{equation}
    \mathbbm{P}\left(\left.x_i^{(t+1)}\neq x_i^{(t)}\right|y_a^{(t)}\right)=\sum_{m:\:y_a^{(t)}+2m-(d-1)<0}\mathbbm{P}\left(Y^{(t)}=m\right)\,,
\end{equation}
with $Y^{(t)}$ a binomial random variable with parameters $d-1$ and $p^{(t)}$.
$Y^{(t)}$ represents the number of satisfied hyperedges in the neighborhood of a node $i$ at time $t$, excluding hyperedge $a$, and $m$ takes all values such that the sum of the values of hyperedges in the neighborhood of $i$ is strictly smaller than zero.

Assuming the initial node values $x_i^{(0)}$ to be independent and identically distributed, $\mathbbm{P}\left(y_a^{(0)}=1\right)$ is the same for all $a\in E$ and can be denoted by $p_0$.
One notes that the update rule \eqref{eq:update-before-mean-field} is then identical for all $a\in E$ and the evolution of the number of satisfied edges in the system can be modeled as:
\begin{align}
    q_{+1}^{(t)}&=\sum_{m<\frac{d}{2}-1}\mathbbm{P}\left(Y^{(t)}=m\right),\;Y^{(t)}\sim\mathcal{B}\left(d-1, p^{(t)}\right)
    \label{eq:mean-field-begin}\,,\\
    q_{-1}^{(t)}&=\sum_{m<\frac{d}{2}}\mathbbm{P}\left(Y^{(t)}=m\right),\;Y^{(t)}\sim\mathcal{B}\left(d-1, p^{(t)}\right)\,,\\
    p^{(t+1)}&=p^{(t)}+\left(1-p^{(t)}\right)\sum_{\mathrm{odd}\;l}\mathbbm{P}\left(X_{-1}^{(t)}=l\right)-p^{(t)}\sum_{\mathrm{odd}\;l}\mathbbm{P}\left(X_{+1}^{(t)}=l\right)\,,
\end{align}
\begin{equation}
    X_{+1}^{(t)}\sim\mathcal{B}\left(k, q_{+1}^{(t)}\right),\;X_{-1}^{(t)}\sim\mathcal{B}\left(k, q_{-1}^{(t)}\right)\,.
    \label{eq:mean-field-end}
\end{equation}
The energy at a given timestep $t$ can then be obtained from $p^{(t)}$ as $e^{(t)}=1-2p^{(t)}$.

The mean field approximation is expected to produce very accurate results for the first few timesteps of the dynamical process, when neighboring hyperedge values have little correlation.
As the system evolves, structures of hyperedge values can emerge, causing the energy to diverge from the mean-field prediction.

The mean field predictions for the energy evolution presented in Fig.~\ref{fig:DCM} are obtained by iteratively computing $p^{(t+1)}$ using \eqref{eq:mean-field-begin}-\eqref{eq:mean-field-end}.
The algorithm is initialized with $p_0=0.5$, except for the case $d=3$, $k=2$ where $p=0.5$ is an unstable fixed point of the equations and $p_0=0.51$ is used instead.
\section{1-Step Replica Symmetry Breaking}\label{app:RSB}
\begin{table}
    \centering
    \begin{tabular}{c||c|c|c|c|c}
\toprule
\midrule
$\boldsymbol{k}$ & $\hspace{20pt}\boldsymbol{2} \hspace{20pt}$ & $\hspace{20pt}\boldsymbol{3} \hspace{20pt}$ & $\hspace{20pt}\boldsymbol{4} \hspace{20pt}$ & $\hspace{20pt}\boldsymbol{5} \hspace{20pt}$ & $\hspace{20pt}\boldsymbol{6} \hspace{20pt}$  \\
\midrule
1dRSB energy & -0.58& -0.87& -0.92& -0.95& -0.96 \\
\midrule
\bottomrule
\end{tabular}
    \caption{\textbf{1dRSB energy for the $k$-XOR-SAT with degree $d=4$}.
The energy at which the transition between the replica symmetric (RS) phase and the 1-step replica symmetry breaking (1dRSB).
The 1dRSB results are obtained from the population dynamics algorithm introduced in~\cite{1RSB-first} applied on \eqref{eq:1RSB-beginning}-\eqref{eq:1RSB-end}.
The phase transition corresponds to the energy at which the RS and 1dRSB results diverge from each other.}
    \label{app::tab:1dRSB}
\end{table}
Equations \eqref{eq:BP-chi-2}-\eqref{eq:free-entropy-2} are written under the \textit{Replica Symmetric} (RS) assumption \cite{1RSB-explanation}.
In the RS phase, almost all trajectories with non-zero measure \eqref{eq:prob-backtrack} belong to the same pure state.
A pure state is defined by a solution $(\chi^\rightarrow,\psi^\rightarrow)$ to the BP equations.
However, this assumption does not hold in general for the low-temperature phase of disordered systems \cite{1RSB-explanation}.
The phase transition from RS to the so-called \textit{Dynamic 1-step Replica Symmetry Breaking} (d1RSB) is usually associated with a transition towards a phase that is hard to sample uniformly~\cite{krzakala2007gibbs}.
The d1RSB phase is characterized by the existence of exponentially many pure states (w.r.t. the system's size $n$) among which the free entropy density is split.

In this section, we describe the equations and algorithm used to find the energy at which the phase transition from RS to d1RSB occurs for the $k$-XOR-SAT model.
To this end, we set $c=0$ and $p=1$, which brings us back to the usual cavity method where the probability distribution \eqref{eq:prob-backtrack} is written over configurations $\mathbf{x}$ instead of trajectories $\underline{\mathbf{x}}$.

We define a probability measure over the pure states as
\begin{equation}
    \mu\left(\chi^\rightarrow,\psi^\rightarrow\right)=\frac{1}{Z_\mu} \exp{\left(n\phi_{int}\left(\chi^\rightarrow,\psi^\rightarrow\right)\right)}\,,
\end{equation}
where $\phi_{int}\left(\chi^\rightarrow,\psi^\rightarrow\right)$ is the \textit{free entropy density} of the pure state defined by the solution $(\chi^\rightarrow,\psi^\rightarrow)$.
The \textit{replicated free entropy density} $\Phi^{\text{d1RSB}}=\frac{1}{n}\log{Z_\mu}$ is then given by
\begin{equation}
    e^{n\Phi^{\text{d1RSB}}}=\sum_{(\chi^\rightarrow,\psi^\rightarrow)}e^{n\phi_{int}(\chi^\rightarrow,\psi^\rightarrow)}\,.
\end{equation}

The replicated free entropy density can be estimated by defining an auxiliary graphical model and applying the BP procedure.
This approach was first introduced in~\cite{1RSB-first}.
Detailed explanations about how to write the BP equations of the auxiliary model can be found in~\cite{1RSB-first,1RSB-explanation,1RSB-explanation-2}.
This results in the following system of self-consistent equations:
\begin{align}
\mathcal{P}_\chi\left(\chi^\rightarrow\right)&\propto\sum_{\psi^\rightarrow}\mathbbm{1}\left\{\chi\rightarrow=\mathcal{F}_\chi\left(\psi^\rightarrow\right)\right\}Z^\chi\left(\psi^\rightarrow\right)\left[\mathcal{P}_\psi\left(\psi^\rightarrow\right)\right]^{k-1}
\label{eq:1RSB-beginning}\,,\\
\mathcal{P}_\psi\left(\psi^\rightarrow\right)&\propto\sum_{\chi^\rightarrow}\mathbbm{1}\left\{\psi\rightarrow=\mathcal{F}_\psi\left(\chi^\rightarrow\right)\right\}Z^\psi\left(\chi^\rightarrow\right)\left[\mathcal{P}_\chi\left(\chi^\rightarrow\right)\right]^{d-1}\,,
\end{align}
\begin{align}
    Z^{\rm edge}_{\text{d1RSB}}&=\sum_{\psi^\rightarrow}Z^{\rm edge}\left(\psi^\rightarrow\right)\left[\mathcal{P}_\psi\left(\psi^\rightarrow\right)\right]^{k}\,,\\
    Z^{\rm node}_{\text{d1RSB}}&=\sum_{\chi^\rightarrow}Z^{\rm node}\left(\chi^\rightarrow\right)\left[\mathcal{P}_\chi\left(\chi^\rightarrow\right)\right]^{d}\,,\\
    Z^{\rm norm}_{\text{d1RSB}}&=\sum_{(\chi^\rightarrow,\psi^\rightarrow)}Z^{\rm norm}(\chi^\rightarrow,\psi^\rightarrow)\mathcal{P}_\chi\left(\chi^\rightarrow\right)\mathcal{P}_\psi\left(\psi^\rightarrow\right)\,,\\
    \Phi^{\text{d1RSB}}&=\frac{d}{k}\log{\left(Z^{\rm edge}_{\text{d1RSB}}\right)}+\log{\left(Z^{\rm node}_{\text{d1RSB}}\right)}-d\log{\left(Z^{\rm norm}_{\text{d1RSB}}\right)}\,.
    \label{eq:1RSB-end}
\end{align}
The phase transition from RS to d1RSB occurs at the point from which the results given by the 1RSB equations \eqref{eq:1RSB-beginning}-\eqref{eq:1RSB-end} diverge from the equations written under the RS assumption, \eqref{eq:BP-chi-2}-\eqref{eq:BP-chi-2}.
Solving the d1RSB equations can be done by means of a population dynamics algorithm explained in~\cite{1RSB-first,koller2024counting}.
The results for the energy at which the d1RSB phase transition occurs for the $k$-XOR-SAT with degree $d=4$ are shown in Table~\ref{app::tab:1dRSB}, and compared in Fig.~\ref{fig:varying-k} with the energies reached by the quench dynamics.

\section{Additional Results}
\subsection{Transient lengths}\label{app:transient}

\begin{figure}%
    \centering
    \begin{subfigure}{.5\textwidth}
        \includegraphics[width=220pt]{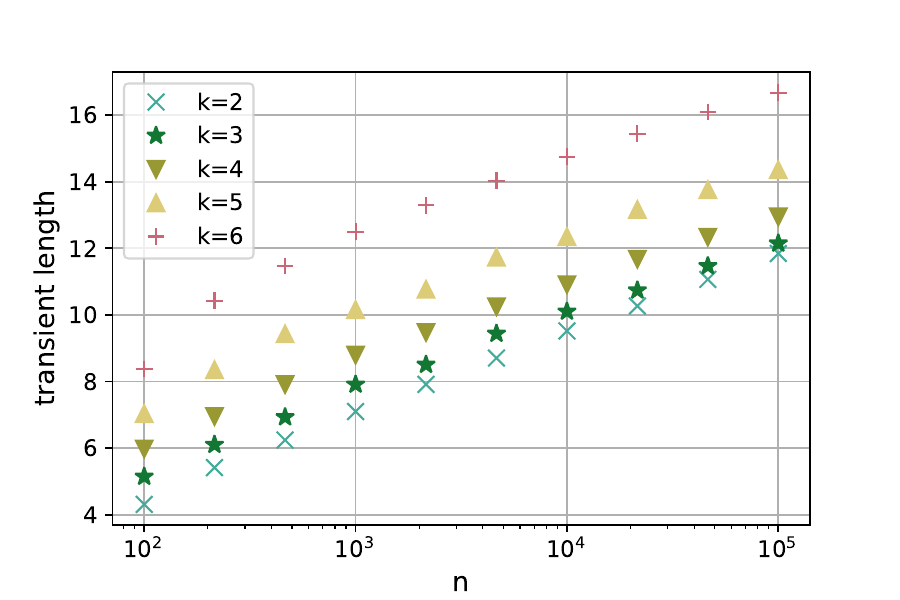}
        \caption{$d=4$}
        \label{fig:transient-even}
    \end{subfigure}\hfill
    \begin{subfigure}{.5\textwidth}
        \includegraphics[width=220pt]{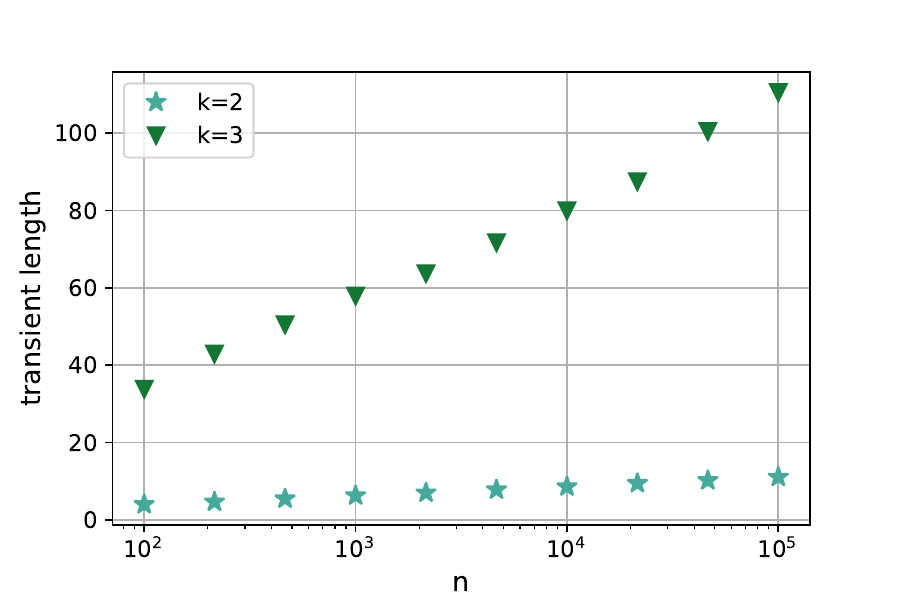}
        \caption{$d=3$}
        \label{app::fig:transient-odd-23}
    \end{subfigure}%
    \begin{subfigure}{.5\textwidth}
        \includegraphics[width=220pt]{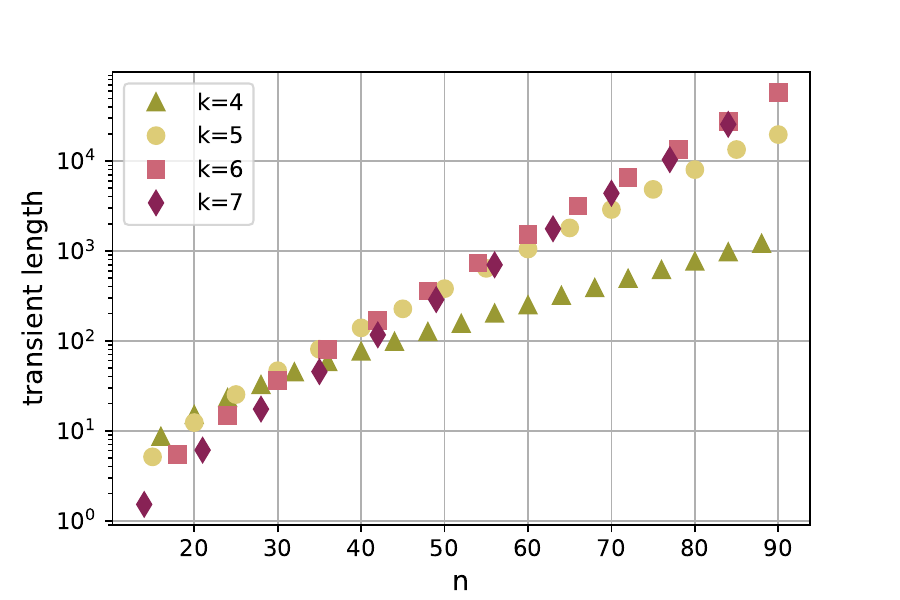}
        \caption{$d=3$}
        \label{app::fig:transient-odd-4}
    \end{subfigure}
    
    \caption{\textbf{Transient lengths of trajectories on hypergraphs with varying orders $k$}.
Numerical results of the transient length as a function of the graph size $n$ for (a) $d=4$ and varying $k$ values, (b) $d=3$ and $k=2,3$, (c) $d=3$ and $k>3$ with smaller values of $n$.
Each result is averaged over $700$ simulations initialized at random with zero magnetization and all simulations were run until convergence to a limit cycle.}\label{app::fig:transient-odd}
\end{figure}

The average transient length of trajectories are shown in Fig.~\ref{fig:transient-even} for hypergraphs with degree $d=4$, and Fig.~\ref{app::fig:transient-odd-23} and \ref{app::fig:transient-odd-4} for hypergraphs with $d=3$, and different values of the order $k$.
In the case $d=4$, the transient length grows logarithmically with the size of the hypergraph $n$, for all considered values of $k$.
The plot indicates that all $k$ values have a similar slope.
However, the offset difference between them increases with $k$.
For $k=3$ on the other hand, only $k=2,3$ are observed to have a logarithmic growth.
While the slope of $k=2$ is of the same order as the $d=4$ case, $k=3$ grows at a significantly greater rate.
For the other observed values $k=4,\ldots,7$, the transient lengths are proportional to the exponential of $n$.
This makes the dynamics on $3$-regular hypergraphs with order $k>3$ very computationally costly to simulate for large values of $n$ or to analyze with methods whose complexities grow exponentially with the length of the considered paths, such as the BDCM.

\subsection{Even degree}\label{app:additional-even}
\begin{table}
\begin{subtable}{1\textwidth}
\centering
    \begin{tabular}{l|rr|rr|rr|rr|rr}
        \toprule\midrule
        $k$ & \multicolumn{2}{l|}{2} & \multicolumn{2}{l|}{3} & \multicolumn{2}{l|}{4} & \multicolumn{2}{l|}{5} & \multicolumn{2}{l}{6} \\
          &             energy & entropy & energy & entropy & energy & entropy & energy & entropy & energy & entropy \\
        $p$ &                    &         &        &         &        &         &        &         &        &         \\
        \midrule
        0 &             -0.362 &   0.603 & -0.417 &   0.677 & -0.458 &   0.724 & -0.489 &   0.758 & -0.514 &   0.783 \\
        1 &             -0.476 &   0.868 & -0.538 &   0.851 & -0.577 &   0.853 & -0.603 &   0.859 & -0.623 &   0.866 \\
        2 &             -0.516 &   0.952 & -0.586 &   0.928 & -0.625 &   0.918 & -0.651 &   0.914 & -0.671 &   0.912 \\
        3 &             -0.533 &   0.981 & -0.613 &   0.967 & -0.653 &   0.955 & -0.679 &   0.946 & -0.698 &   0.940 \\
        4 &             -0.541 &   0.992 & -0.629 &   0.985 & -0.672 &   0.976 & -0.698 &   0.967 & -0.717 &   0.959 \\
        5 &             -0.545 &   0.997 & -0.639 &   0.994 & -0.686 &   0.988 & -0.713 &   0.980 & -0.731 &   0.972 \\
        6 &             -0.546 &   0.999 & -0.644 &   0.998 & -0.695 &   0.994 & -0.723 &   0.989 & -0.742 &   0.981 \\
        7 &             -0.547 &   0.9995 & -0.647 &   0.999 & -0.701 &   0.998 & -0.732 &   0.994 &    - &     - \\
        \midrule\bottomrule
    \end{tabular}
    \caption{$d=4$}
    \end{subtable}
    
    \begin{subtable}{1\linewidth}
    \vspace{.5cm}
        \resizebox{\columnwidth}{!}{%
\begin{tabular}{l|cc|cc|cc|cc|cc|cc|cc|cc}
\toprule
\midrule
p & \multicolumn{2}{l}{0} & \multicolumn{2}{l}{1} & \multicolumn{2}{l}{2} & \multicolumn{2}{l}{3} & \multicolumn{2}{l}{4} & \multicolumn{2}{l}{5} & \multicolumn{2}{l}{6} & \multicolumn{2}{l}{7} \\
  & energy & entropy & energy & entropy & energy & entropy & energy & entropy & energy & entropy & energy & entropy & energy & entropy & energy & entropy \\
d &  &  &  &  &  &  &  &  &  &  &  &  &  &  &  &  \\
\midrule
4 & -0.417 & 0.677 & -0.538 & 0.851 & -0.586 & 0.928 & -0.613 & 0.967 & -0.629 & 0.985 & -0.639 & 0.994 & -0.644 & 0.998 & -0.647 & 0.999 \\
6 & -0.361 & 0.637 & -0.464 & 0.807 & -0.506 & 0.890 & -0.530 & 0.936 & -0.544 & 0.963 & -0.555 & 0.979 & -0.562 & 0.989 & - & - \\
8 & -0.324 & 0.612 & -0.415 & 0.778 & -0.453 & 0.863 & -0.474 & 0.913 & -0.487 & 0.944 & -0.497 & 0.964 & - & - & - & - \\
10 & -0.297 & 0.594 & -0.379 & 0.758 & -0.414 & 0.844 & -0.433 & 0.896 & -0.446 & 0.929 & -0.454 & 0.951 & - & - & - & - \\
12 & -0.276 & 0.581 & -0.352 & 0.742 & -0.384 & 0.829 & -0.402 & 0.882 & -0.413 & 0.916 & -0.421 & 0.940 & - & - & - & - \\
14 & -0.259 & 0.570 & -0.330 & 0.730 & -0.360 & 0.817 & -0.377 & 0.870 & -0.388 & 0.905 & - & - & - & - & - & - \\
16 & -0.245 & 0.561 & -0.312 & 0.720 & -0.340 & 0.806 & -0.356 & 0.860 & -0.366 & 0.896 & - & - & - & - & - & - \\
18 & -0.233 & 0.554 & -0.318 & 0.535 & -0.324 & 0.798 & -0.339 & 0.852 & -0.348 & 0.889 & - & - & - & - & - & - \\
20 & -0.223 & 0.548 & -0.283 & 0.705 & -0.309 & 0.791 & -0.323 & 0.845 & - & - & - & - & - & - & - & - \\
22 & -0.214 & 0.542 & -0.271 & 0.698 & -0.296 & 0.784 & -0.310 & 0.839 & - & - & - & - & - & - & - & - \\
\midrule
\bottomrule
\end{tabular}
}
\caption{$k=3$}
    \end{subtable}
    \caption{\textbf{BDCM energies and entropies for quench dynamics on $d$-regular $k$-uniform hypergraphs with even $d$}.
Entropy obtained from the BP messages, after convergence of the BDCM iterating process, as well as the corresponding energy of a typical attractor, for attractors of length $c=1$ and different values of incoming path lengths $p$, the degree $d$ and the order $k$ of the hypergraph. (a)~Fixed degree $d=4$ and varying values of the order $k$. (b)~Fixed order $k=3$ and varying even values of the degree $d$.}
    \label{app::tab:BDCM-even-degree-evenk-stay}
\end{table}
Table~\ref{app::tab:BDCM-even-degree-evenk-stay} contains all BDCM results for the entropy and energy of backtracking attractors with incoming path length $p=1,\ldots,7$ on $4$-regular hypergraphs with order $k=2,\ldots,6$, and $k$-uniform hypergraphs with even degree values in $4,\ldots,22$.
They are obtained using \eqref{eq:saddle-point} and \eqref{eq:obtain-observable} respectively after convergence of the BP messages \eqref{eq:BP-chi-2} and \eqref{eq:BP-psi-2}. 
\begin{figure}
    \centering
    \begin{subfigure}{.5\textwidth}
        \centering
        \includegraphics[width=231pt]{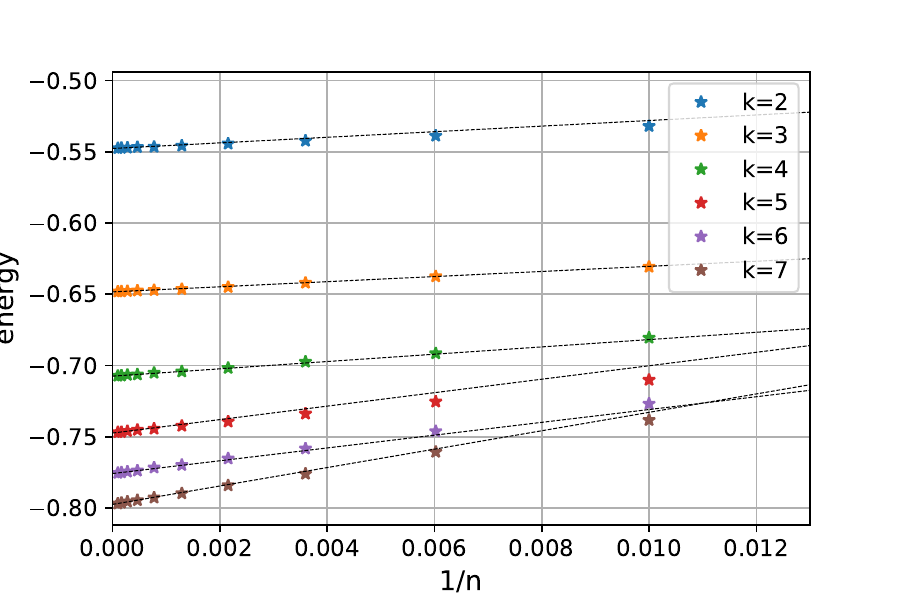}
        \label{fig:aaa}
        \caption{}
        \label{app::fig:convergence-energy-numerics}
    \end{subfigure}%
    \begin{subfigure}{.5\textwidth}
        \centering
        \includegraphics[width=220pt]{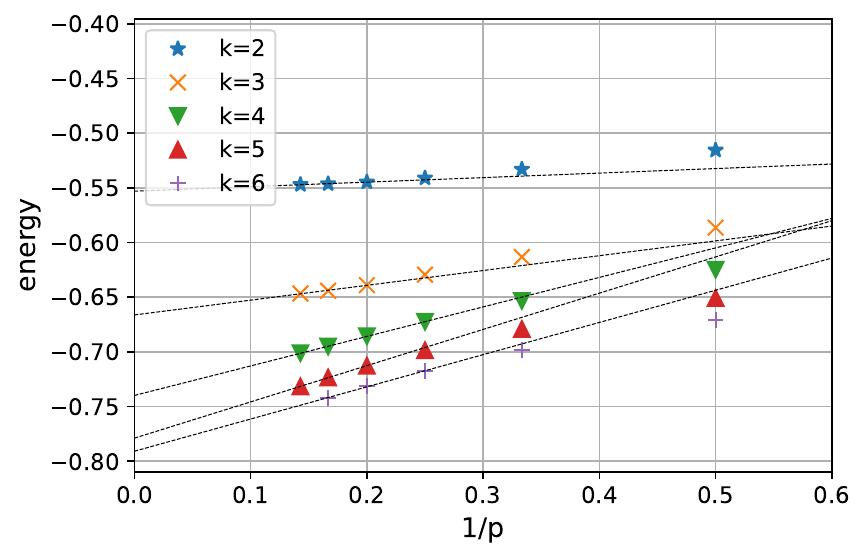}
        \caption{}
        \label{app::fig:convergence-energy-BDCM}
    \end{subfigure}
    \caption{\textbf{Extrapolation of the energy in the attractor on $4$-regular $k$-uniform hypergraphs}.
(a) Numerical result for the energy in the attractor versus the inverse graph size $\frac{1}{n}$.
Each result is obtained by averaging over $7000$ simulations initialized at random with zero magnetization.
A linear fit is shown for each value of $k$ (b) BDCM results for the energy $e_{\attr}$ of a typical attractor of length $c=1$ as a function of the inverse incoming path length $\frac{1}{p}$, a linear fit of the 3 last points is shown for each value of $k$.
The energy values extrapolated at $\frac{1}{n}=0$ and $\frac{1}{p}=0$ respectively are used on Fig.~\ref{fig:BDCM-energies}.}
    \label{app::fig:convergence-energy}
\end{figure}
The results for the energy are then used in Fig.~\ref{app::fig:convergence-energy-BDCM} to obtain an approximation of the energy in the attractor of a typical trajectory.
The last $3$ points for each $k$ are linearly fitted and extrapolated at $\frac{1}{p}=0$.
The same procedure is applied in Fig.~\ref{app::fig:convergence-energy-numerics} to obtain the extrapolation of the numerical results for the energies in the limit $n\rightarrow\infty$.
Those values are used in Fig.~\ref{fig:varying-k} and compared with the ground state energies of the system and the energies under which the replica symmetric ansatz does not hold anymore.
\subsection{Odd degree}\label{app:additional-odd}
\begin{figure}
    \centering
    \begin{minipage}[t]{.5\linewidth}
    \centering
    \vspace{0pt}
        \includegraphics[width=240pt]{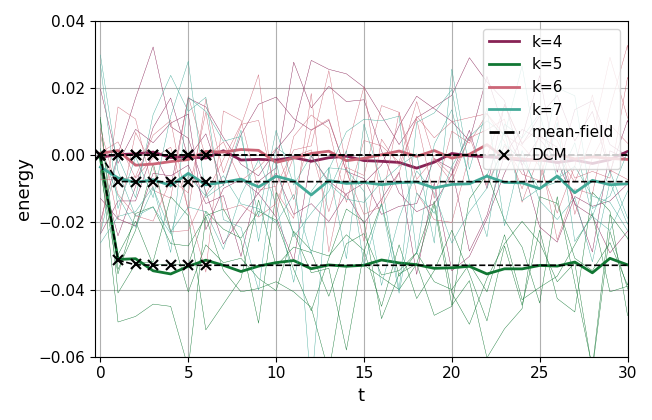}
        \caption*{(a)}
    \end{minipage}%
    \begin{minipage}[t]{.5\linewidth}
    \centering
    \vspace{15pt}
        \resizebox{0.9\columnwidth}{!}{%
        \begin{tabular}{l|cc|cc}
        \toprule
        \midrule
        k & \multicolumn{2}{l|}{5} & \multicolumn{2}{l}{7} \\
          & DCM & mean field & DCM & mean field \\
        p &  &  &  &  \\
        \midrule
        0 & 0.000000 & 0.000000 & 0.000000 & 0.000000 \\
        1 & -0.031250 & -0.031250 & -0.007812 & -0.007812 \\
        2 & -0.032473 & -0.032624 & -0.007853 & -0.007856 \\
        3 & -0.032694 & -0.032747 & -0.007853 & -0.007856 \\
        4 & -0.032724 & -0.032758 & -0.007853 & -0.007856 \\
        5 & -0.032728 & -0.032759 & -0.007853 & -0.007856 \\
        6 & -0.032729 & -0.032759 & -0.007853 & -0.007856 \\
        \midrule
        \bottomrule
        \end{tabular}
        }
        \vspace{20pt}
        \caption*{(b)}
    \end{minipage}
    \caption{\textbf{Evolution of the energy from numerical simulations compared to the dynamical cavity method and the mean-field model on $3$-regular $k$-uniform hypergraphs, with $k\geq4$}.
(a) 5 simulated trajectories of the energy of the system, initialized at random with zero magnetization, are shown in thin lines.
The average energy taken over $100$ trajectories is displayed with bold lines. The energies obtained from the DCM with $p=6$ and the mean-field energy computed by iterating equations~\eqref{eq:mean-field-begin}-\eqref{eq:mean-field-end}, initialized with $p_0=0.5$ are displayed on (a) and reported in table~(b) for $k=5,7$. For even values of $k$, all founded results of the DCM are 0.}
    \label{app::fig:mean-field-odd-4}
\end{figure}
\begin{figure}
    \centering
        \includegraphics[width=250pt]{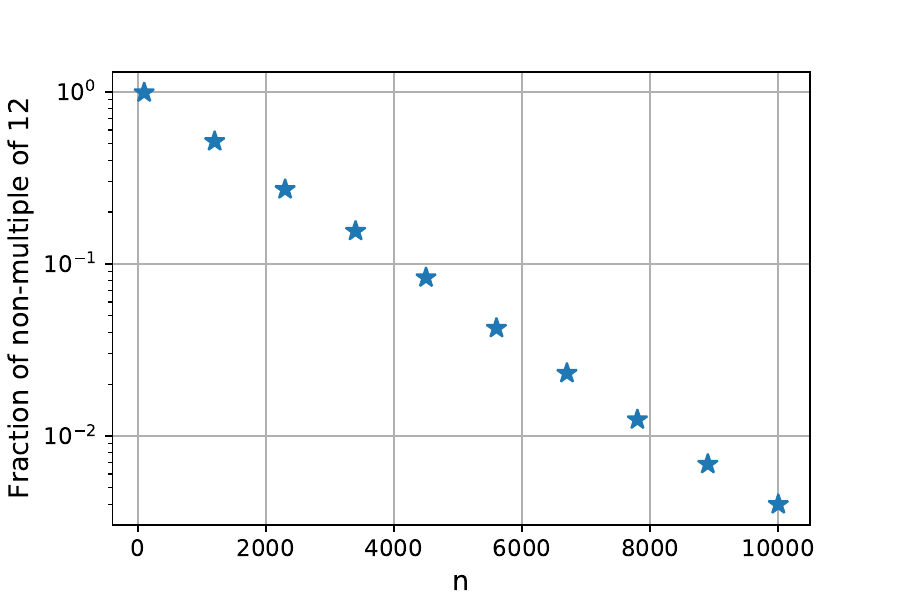}
    \caption{\textbf{Fraction of trajectories on $3$-regular $3$-uniform hypergraphs that does not end up in limit cycles whose length is a multiple of $12$.} The fraction of observed cycle lengths that are multiples of $12$ are plotted as a function of the hypergraph size $n$.
Each fraction was taken over $7000$ numerical simulations initialized at random with zero magnetization and all simulations were run until convergence to a limit cycle.}
    \label{app::fig:12-multiples}
\end{figure}

\paragraph{Evolution of the energy for $k>3$}Fig.~\ref{app::fig:mean-field-odd-4} shows the numerical evolution of the energy $e^{(t)}$ for the quench dynamics on $3$-regular hypergraphs with $k=4,\ldots,7$ and compares it to the mean field prediction.
For each $k=4,\ldots,7$, the mean-field equations have a stable fixed point between $-0.1$ and $0$.
The mean field energies initialized with $p_0=0.5$ are observed to quickly approach their respective stable fixed points.
We see that the averages of the numerical energies closely follow the mean-field predictions.
The numerical trajectories chaotically oscillate around the mean-field fixed points and are unable to reach low energies.
\paragraph{Cycle lengths for $d=3$ and $k=3$ are multiples of $12$}
In Fig.~\ref{app::fig:12-multiples}, we see that the fraction of trajectories that does not end up in an attractor whose length is a multiple of $12$ decreases at an exponential rate as the hypergraph size $n$ increases.
This indicates that in the large system limit, almost all trajectories ruled by this process eventually converge to cycles with length $c=12X(n)$ for some random variable $X$ whose (unknown) distribution depends on the hypergraph size $n$.

Table~\ref{app::tab:BDCM-odd} contains the BDCM results for the entropy and energy of backtracking attractors with incoming path length $p$ on $4$-regular hypergraphs with order $k=2,\ldots,6$.
The entropy and energy are computed for different cycle lengths $c=1,2,4$ and all values of $p$ for which the BDCM algorithm is computationally tractable.
The entropy results for $k=3$ are plotted in Fig.~\ref{fig:entropy-k3}.
\begin{table}
    \centering
    \begin{tabular}{ll|rr|rr}
\toprule
\midrule
  & $k$ & \multicolumn{2}{l|}{2} & \multicolumn{2}{l}{3} \\
  &   &             energy & entropy & energy & entropy \\
$c$ & $p$ &                    &         &        &         \\
\toprule
1 & 0 &             -0.600 &   0.339 & -0.660 &   0.493 \\
  & 1 &             -0.688 &   0.587 & -0.745 &   0.627 \\
  & 2 &             -0.724 &   0.676 & -0.770 &   0.705 \\
  & 3 &             -0.745 &   0.709 & -0.781 &   0.757 \\
  & 4 &             -0.758 &   0.725 & -0.801 &   0.792 \\
  & 5 &             -0.766 &   0.732 & -0.815 &   0.817 \\
\midrule
2 & 0 &              0.000 &   0.678 & -0.500 &   0.595 \\
  & 1 &              0.000 &   0.903 & -0.563 &   0.713 \\
  & 2 &              0.000 &   0.969 & -0.598 &   0.779 \\
  & 3 &              0.000 &   0.988 & -0.611 &   0.826 \\
  & 4 &              0.000 &   0.996 & -0.627 &   0.857 \\
  & 5 &              0.000 &   0.998 & -0.639 &   0.879 \\
\midrule
4 & 0 &              0.000 &   0.678 & -0.494 &   0.601 \\
  & 1 &              0.000 &   0.903 & -0.552 &   0.719 \\
  & 2 &              0.000 &   0.969 & -0.586 &   0.785 \\
  & 3 &              0.000 &   0.988 & -0.598 &   0.831 \\
\midrule
\bottomrule
\end{tabular}
    \caption{\textbf{BDCM energies and entropies for quench dynamics on $3$-regular $k$-uniform hypergraphs}.
Entropy obtained from the BP messages, after convergence of the BDCM iterating process, as well as the corresponding energy of a typical attractor, for attractors of length $c=1,2,4$ on hypergraphs with order $k=2,3$, and different values of incoming path lengths $p$.}
    \label{app::tab:BDCM-odd}
\end{table}
\end{appendices}
\end{document}